%% file: arttotal.tex
\documentstyle[12pt]{article}

\renewcommand{\S}{S_{\mbox{\scriptsize vac}}}
\newcommand{\s}{S_{\mbox{\scriptsize cl}}}
\newcommand{\T}{T_{\mbox{\scriptsize vac}}}
\renewcommand{\t}{T_{\mbox{\scriptsize cl}}}
\newcommand{\E}{{\cal E}_{\mbox{\scriptsize vac}}}
\newcommand{\e}{{\cal E}_{\mbox{\scriptsize cl}}}
\renewcommand{\P}{\hat P}
\newcommand{\J}{\hat J}
\newcommand{\D}{\hat D}
\renewcommand{\d}{{\hat D}_{\bf 1}}
\newcommand{\C}{\hat C}
\newcommand{\R}{\hat{\cal R}}
\newcommand{\I}{{\cal I}^+}
\renewcommand{\|}{\biggl |_{{\cal I}^+}}
\newcommand{\1}{\biggl |_{\textstyle \mbox{path}\; 1}}
\newcommand{\2}{\biggl |_{\textstyle \mbox{path}\; 2}}
\newcommand{\3}{\biggl |_{\textstyle \mbox{path}\; 3}}
\newcommand{\tr}{{\mbox{tr}}}
\newcommand{\integral}{\int dx\, g^{1/2}}
\newcommand{\sphere}{\int d^2 {\cal S} (\phi)\,}
\newcommand{\cint}{
\int\limits_{\mbox{\scriptsize past of}\; x} d{\bar x}\,{\bar g}^{1/2}
}
\makeatletter
\@addtoreset{equation}{section}
\makeatother
\begin{document}
\input arttxt01.tex
\newpage
\input arttxt02.tex

\newpage
\input arttxt03.tex

\newpage
\input arttxt04.tex

\newpage
\input arttxt05.tex

\newpage
\input arttxt06.tex

\newpage
\input arttxt07.tex
\newpage
\input arttxt08.tex

\newpage
\input arttxt09.tex

\newpage
\appendix
\renewcommand{\thesection}{Appendix\enskip\Alph{section}.}

\input arttxt10.tex
\newpage

\input arttxt11.tex

\newpage

\input arttxt12.tex
\newpage

\input arttxt13.tex

\newpage
\input arttxt14.tex

\newpage
\input figures.tex

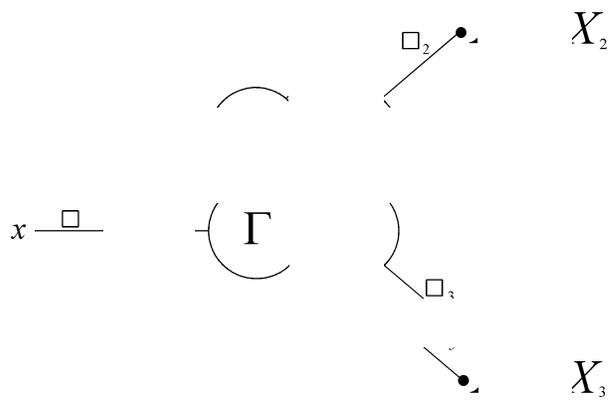
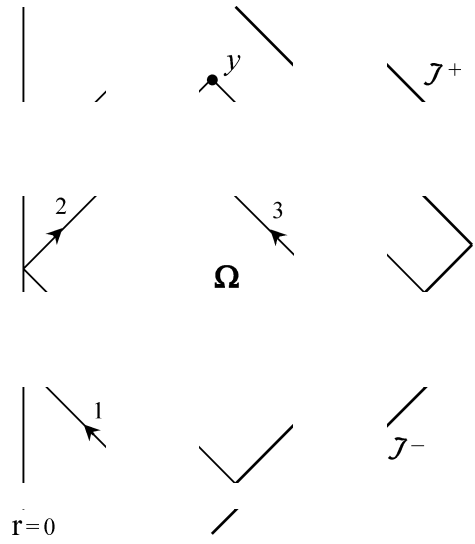
\end{document}

%% file: arttxt01.tex
{\renewcommand{\theequation}{1.\arabic{equation}}

\begin{center}
{\LARGE\bf 
Vacuum amplification}\\{\LARGE\bf
of the high - frequency}\\{\LARGE\bf 
electromagnetic radiation}\\
\end{center}
\begin{center}
{\bf G.A. Vilkovisky}
\end{center}
\begin{center}
Lebedev Physics Institute and
Research Center in Physics,\\ Leninsky Prospect 53, Moscow 117924,
Russia \\
\end{center}
\vspace{2cm}
\hspace{7cm}{\large\it In memory of David Kirzhnits}
\vspace{2cm}
\begin{abstract}

When an electrically charged source is capable of both emitting
the electromagnetic waves and creating charged particles from
the vacuum, its radiation gets so much amplified that only the
backreaction of the vacuum makes it finite. The released energy 
and charge are calculated in the high-frequency approximation.
The technique of expectation values is advanced and employed.
\end{abstract}
\newpage

\begin{center}
\section{\bf    Introduction and summary} 
\end{center}

$$ $$ 

Reaction of the vacuum on rapidly moving sources, or strongly variable
fields is important for the evolution of black holes and early
universe but is interesting also in electrodynamics. We know that,
in electrodynamics, the vacuum attenuates an external charge. 
Suppose now that the external source is not a monopole but, say,
a dipole, and let this dipole be capable of emitting the electromagnetic
waves so that the information about it reaches infinity. Then what
will be the vacuum effect on such a dipole ?

The answer obtained below is that the effect is opposite; a radiation
of the dipole gets amplified. This effect becomes  noticeable as soon
as the typical frequency of the dipole exceeds the threshold of pair
creation. A flux of charged particles that appears in this case is
accompanied by an {\it increase} of the electromagnetic radiation.
Generally, there is a nonlocal tail of radiation caused by the vacuum
stress but, at a high frequency, the effect boils down to a multiplication
of the classical radiation rate by a renormalization constant. Since
the dipole is a nonlocal object, its renormalization\footnote{Residual
after an infinite renormalization of the monopole.} is finite and
observable.

The vacuum amplification of the electromagnetic waves emitted by a source
is analogous to the effect of the vacuum gravitational waves [1]. The
difference is only in the theoretical mechanisms and in the dimensions 
of the coupling constants\footnote{The main difference is, of course,
in the fact that, for a creation of the gravitational charge, there is
no threshold.}. The dimension of the coupling constant causes that the
gravitational effect never boils down to a mere renormalization.

In the case of electromagnetic waves, the mechanism by which this effect
emerges in theory is as follows. If one calculates the energy of charged
particles created from the vacuum by a given nonstationary electromagnetic
field [2], one finds that the result can be obtained only in the case
where the electromagnetic field contains no outgoing waves. In the general
case this energy is infrared divergent with the divergent term proportional
to the energy of the outgoing waves. The appearing divergence is a signal
that the calculation is not complete because {\it the energy of the vacuum
of charged particles goes partially into the coherent electromagnetic
radiation.} Indeed, the missing contribution comes from the backreaction
of the vacuum on the electromagnetic field. If one calculates the
effective electromagnetic field that solves the expectation-value
equations, one finds that the quantum correction to the energy of
the outgoing waves is also divergent, and the two divergences cancel
each other. As a result, the total released energy is finite but is no
more the energy of created charged particles alone. Rather it is a sum
of the energy that goes with charged particles and the energy that goes
with the enhanced electromagnetic waves. The two contributions can then
be separated by calculating the released charge.

It is worth noting that the vacuum reactions on the low-frequency and
high-frequency external fields are very different. The effects like
the anomalous magnetic moment in QED refer to the low-frequency 
electromagnetic fields and are not related to the effect considered
here. On the other hand, in the mechanism described above one easily
recognizes the physics that stands behind the so-called infrared
disaster. Here this physics actually works, and, of course, there
is no disaster if one considers not the transitions between concocted
states but the evolution of expectation values.

The terms "high-frequency approximation" and "strongly variable field"
are used here as synonyms. Let $l$ be the typical spatial size of the
source of an external field and $\nu$ be its typical frequency. On the
other hand, let $m$ be the mass of the lightest particles interacting
with this field. In the problem of the vacuum particle creation, the
external field is considered as strongly variable if the energy
$\hbar\nu$ dominates both the rest energy of the vacuum particle and
its static energy in this field. The first of these conditions is
discussed in Sec. 4 below, and its more accurate form is
\begin{equation}
\hbar\nu\gg mc^2\left(\frac{mc}{\hbar}l\right)\quad.
\end{equation}
The second is exemplified in [2]. Under condition (1.1) the vacuum
particles may be regarded as massless in the calculation of their
fluxes. However, the mass $m$ cannot be neglected in the calculation
of the static polarization and charge renormalization.

It makes sense to begin with quoting the result for the energy of 
particles  created from the vacuum by strongly variable fields
of arbitrary configurations. The respective calculation was carried
out in [2] for {\it the standard loop}, i.e. for the vacuum action 
of the form
\begin{equation}
\S=\frac{\mbox{i}}{2}\log\mbox{det}{\hat H}\quad,
\end{equation}
\begin{equation}
{\hat H}=g^{\mu\nu}\nabla_{\mu}\nabla_{\nu}{\hat 1}+
\left(\P-\frac{1}{6}R{\hat 1}\right)-m^2{\hat 1}
\end{equation}
where the operator ${\hat H}$ is defined as acting on an arbitrary
set of quantum fields. The hat over a symbol means that this symbol
is a matrix in the space of field components, ${\hat 1}$ is the unit
matrix, and the matrix trace will be denoted $\tr$. The external fields
in (1.3) are the metric $g_{\mu\nu}$, the matrix potential $\P$, and
an arbitrary connection defining the commutator curvature:
\begin{equation}
[\nabla_{\mu},\nabla_{\nu}]=\R_{\mu\nu}\quad.
\end{equation}
The sign convention for the Ricci scalar $R$ in (1.3) is such that,
when acting on a scalar field, the operator ${\hat H}$ with $\P=0$
and $m=0$ is conformal invariant.

In the present paper, only the effect of the commutator curvature
$\R_{\mu\nu}$ is considered, and {\it nevertheless} the action (1.2)
is needed with all the three types of external fields present.
The dependence of $\S$ on the metric is needed because the vacuum
energy-momentum tensor is obtained by varying the action (1.2)
with respect to the metric
\begin{equation}
\T^{\mu\nu}=\frac{2}{g^{1/2}}\frac{\delta\S}{\delta g_{\mu\nu}}
\end{equation}
(and next using the retarded resolvent for the nonlocal form factors
\footnote{See [2] and references therein.}). The dependence of
$\S$ on the potential is needed because the results for various
quantum field models, e.g. for the spinor QED, are obtained
by combining the standard loops with $\P$ generally depending
on $\R_{\mu\nu}$ (see [3] and Sec. 8 below).

For the classical action of the commutator curvature one may take
the expression
\begin{equation}
\s=\frac{1}{16\pi\kappa^2}\integral\tr\; \R_{\mu\nu}\R^{\mu\nu}
\end{equation}
with some coupling constant $\kappa^2>0$. In the case of the 
electromagnetic connection, the $\kappa^2$ is to be chosen so
that (1.6) be the Maxwell action
\begin{equation}
\s=-\frac{1}{16\pi}\integral F_{\mu\nu}F^{\mu\nu}
\end{equation}
(the matrix trace in (1.6) is always negative [2]). Denote
\begin{equation}
\t^{\mu\nu}=\frac{2}{g^{1/2}}\frac{\delta\s}{\delta g_{\mu\nu}}\quad .
\end{equation}

The energy of the classical electromagnetic radiation and, in the
high-frequency approximation, also the outgoing flux of the vacuum
energy can be calculated at the future null infinity $\I$ [4,2].
The limit $\I$ is defined as the limit of infinite luminosity
distance $r$ along the null geodesic that, when traced to the future,
comes at the instant $u$ of retarded time to the point $\phi$ of the
celestial 2-sphere $\cal S$. One has
\begin{eqnarray}
\frac{1}{4}\nabla_{\mu}v\nabla_{\nu}v\;\t^{\mu\nu}\|&=&
-\frac{1}{r^2}\frac{\partial}{\partial u}\,\e(u,\phi)+
O\Bigl(\frac{1}{r^3}\Bigr)\quad,\\
\frac{1}{4}\nabla_{\mu}v\nabla_{\nu}v\;\T^{\mu\nu}\|&=&
-\frac{1}{r^2}\frac{\partial}{\partial u}\,\E(u,\phi)+
O\Bigl(\frac{1}{r^3}\Bigr)
\end{eqnarray}
where
\begin{equation}
\nabla v\|=\nabla u +2\nabla r\quad,\quad (\nabla u,\nabla r)\|=-1\quad.
\end{equation}

The notation $\partial{\cal E}/\partial u$ is introduced to represent
the energy loss. Taken with the sign minus, each $\partial{\cal E}/
\partial u$ is the density of the respective outgoing flux of energy
so that the total emitted energy is obtained by integrating
$\left(-\partial {\cal E}/\partial u\right)$ over the 2-sphere $\cal S$
(normalized to have the area $4\pi$) and the time $u$. Specifically,
the total released vacuum energy equals
\begin{equation}
\int\limits_{-\infty}^{\infty}du\sphere \left(-\frac{\partial\E}{
\partial u} \right)=\sum_A \varepsilon_A \,\langle\mbox{in vac}\;
|a^{+A}_{\mbox{\scriptsize out}}a^{A}_{\mbox{\scriptsize out}}
|\;\mbox{in vac}\rangle
\end{equation}
and equals the total energy of particles created from the in-vacuum
by external fields (see, e.g., [5]). Here
$a^{+A}_{\mbox{\scriptsize out}}$, $a^{A}_{\mbox{\scriptsize out}}$
are the creation and annihilation operators for the out-vacuum, and
$\varepsilon_A$ is the energy in the out-mode $A$. Since $\t^{\mu\nu}$
is energy-dominant [4], the flux $\left(-\partial\e/\partial u\right)$
is manifestly positive. The flux $\left(-\partial\E/\partial u\right)$
is sign-indefinite because of the quantum uncertainty but the integrated
flux (1.12) is positive [2].

Only the external fields generated by sources are considered in [2]
and the present paper. The sources of external fields in (1.3) are
\begin{equation}
\J=\P\quad ,\quad \J^{\mu}=\nabla_{\nu}\R^{\mu\nu}\quad ,\quad
J^{\mu\nu}=R^{\mu\nu}-\frac{1}{2}g^{\mu\nu}R
\end{equation}
where $R^{\mu\nu}$ is the Ricci tensor of the external metric, and
the potential $\P$ is identified with its own source. These classical
sources will be referred to as {\it bare} sources. The bare sources
are assumed to have their supports in a spacetime tube with compact
spatial sections and a timelike boundary. Their domain of nonstationarity
is assumed compact in both space and time [2].

At a large distance from a source, all its manifestations at both
classical and quantum levels are governed by a single quantity,
its {\it radiation moment} [2] defined as an integral of the source
over a spacelike {\it hyperplane}. The hyperplane itself is defined
as follows. One considers all timelike geodesics that, when traced
to the future, reach infinity with one and the same value of energy
per unit rest mass ($E>1$) and at one and the same point of the
celestial sphere ($\phi\in{\cal S}$). These geodesics make a 3-parameter
congruence which is hypersurface-orthogonal, 
and the hyperplanes are the hypersurfaces orthogonal
to this congruence [2]. Let
\begin{equation}
T_{\gamma\phi}(x)=\mbox{const.}
\end{equation}
be the equation of these hypersurfaces. The parameter $\gamma$ that,
along with $\phi$, labels the function $T_{\gamma\phi}(x)$ is a
redefined $E$:
\begin{equation}
\gamma=\frac{\sqrt{E^2-1}}{E}\quad , \quad 0<\gamma<1
\end{equation}
and the function $T_{\gamma\phi}(x)$ itself is normalized by the
condition
\begin{equation}
\left(\nabla T_{\gamma\phi}(x)\right)^2=-(1-\gamma^2)\quad .
\end{equation}
The radiation moments of the sources in (1.13) are the following
integrals [2]:
\begin{eqnarray}
\D&=&\frac{1}{4\pi}\integral \delta\left(T_{\gamma\phi}(x)-u\right)\J(x)
\quad ,\\
\D^{\alpha}&=&\frac{1}{4\pi}\integral \delta\left(T_{\gamma\phi}(x)-u\right)
g^{\alpha}_{\;\;\bar\alpha}\J^{\bar\alpha}(x)\quad ,\\
D^{\alpha\beta}&=&\frac{1}{4\pi}\integral \delta\left(T_{\gamma\phi}(x)-u
\right)g^{\alpha}_{\;\;\bar\alpha}g^{\beta}_{\;\;\bar\beta}J^{{\bar\alpha}
{\bar\beta}}(x) 
\end{eqnarray}
where $g^{\alpha}_{\;\;\bar\alpha}$ is the propagator of the geodetic
parallel transport [6] connecting the integration point with the
future end point of the geodesic having the parameters $\gamma$,
$\phi$. The moments are tensors at this end point depending parametrically
on time $u$.

At the limit $\gamma=1$ the hyperplane (1.14) becomes null. The vector
and tensor moments taken at $\gamma=1$ govern the classical electromagnetic 
and gravitational radiation. Specifically, for the energy of the
electromagnetic waves one has [2]
\begin{equation}
-\frac{\partial}{\partial u}\e(u,\phi)=-\frac{1}{4\pi\kappa^2}\,\tr\;
g_{\alpha\beta}\Bigl(\frac{\partial}{\partial u}\D^{\alpha}\Bigr)
\Bigl(\frac{\partial}{\partial u}\D^{\beta}\Bigr)\biggl|_{\gamma=1}\quad .
\end{equation}
The expansion of the vector and tensor moments at $\gamma=0$ gives rise
to the usual multipole moments [2]. The radiation moments integrated
over $\gamma$ govern the energy of the vacuum particle production.
One has [2]
\begin{eqnarray}
-\frac{\partial}{\partial u}\E(u,\phi)&=&\frac{1}{(4\pi)^2}
\int\limits^1_0 d\gamma \gamma^2\,\tr \left[\Bigl(\frac{\partial^2}{
\partial u^2}\D\Bigr)\Bigl(\frac{\partial^2}{\partial u^2}\D\Bigr)
\right.\nonumber\\
&&{}-\frac{1}{3}\frac{1}{(1-\gamma^2)} g_{\alpha\beta}
\Bigl(\frac{\partial}{\partial u}\D^{\alpha}\Bigr)\Bigl(
\frac{\partial}{\partial u}\D^{\beta}\Bigr)\nonumber\\
&&\left. {}+\frac{{\hat 1}}{30}(g_{\mu\alpha}g_{\nu\beta}-
\frac{1}{3}g_{\mu\nu}g_{\alpha\beta})
\Bigl(\frac{\partial^2}{\partial u^2}D^{\mu\nu}\Bigr)
\Bigl(\frac{\partial^2}{\partial u^2}D^{\alpha\beta}\Bigr)\right]+
\mbox{Q.N.}
\end{eqnarray}
where the abbreviation Q.N. means Quantum Noise and denotes the
sign-indefinite contribution that is present in the vacuum energy
flux because of the quantum uncertainty but sums to zero for the
whole history [2,5]:
\begin{equation}
\int\limits_{-\infty}^{\infty}du\sphere\left(\mbox{Q.N.}\right)=0\quad .
\end{equation}
In the equations below the term Q.N. will often be omitted but its
presence will be tacitly assumed in all expressions for the vacuum
energy.

Expression (1.21) is the starting point of the present work. It is
seen that in the case of the vector moment (and only in this case)
the validity of this expression is limited by the condition
\begin{equation}
\tr\; g_{\alpha\beta}\Bigl(\frac{\partial}{\partial u}\D^{\alpha}\Bigr)
\Bigl(\frac{\partial}{\partial u}\D^{\beta}\Bigr)\biggl|_{\gamma=1}=0
\end{equation}
which in view of (1.20) is a condition that the vector source does not
radiate classically:
\begin{equation}
\frac{\partial}{\partial u}\,\e(u,\phi)\equiv 0\quad .
\end{equation}
If it does, the integral in (1.21) has a pole at $\gamma=1$.

Below I shall consider only the contribution of the vector source
assuming that the other contributions in (1.21) are absent. It is
useful to decompose the vector moment over the vector basis at
infinity:
\begin{equation}
g_{\alpha\beta}\biggl|_{r\to\infty}=-\nabla_{\alpha}u\nabla_{\beta}u
-(\nabla_{\alpha}u\nabla_{\beta}r+\nabla_{\alpha}r\nabla_{\beta}u)+
\frac{1}{2}(m_{\alpha}m^*_{\beta}+m^*_{\alpha}m_{\beta})
\end{equation}
where $m$ is the complex null vector tangent to the 2-sphere ${\cal S}$,
and $m^*$ is its complex conjugate. The projection
\begin{equation}
\Bigl(\nabla_{\mu}u+(1-\gamma)\nabla_{\mu}r\Bigr)\D^{\mu}
\equiv{\hat e}=\mbox{const.}
\end{equation}
is the full conserved charge of the bare source [2]. Hence
\begin{equation}
\tr\;
 g_{\alpha\beta}\Bigl(\frac{\partial}{\partial u}\D^{\alpha}\Bigr)
\Bigl(\frac{\partial}{\partial u}\D^{\beta}\Bigr)=
\tr \left[\Bigl(m_{\alpha}\frac{\partial}{\partial u}\D^{\alpha}\Bigr)
\Bigl(m_{\beta}\frac{\partial}{\partial u}\D^{\beta}\Bigr)^* +
(1-\gamma^2)\Bigl(\nabla_{\alpha}r\frac{\partial}{\partial u}\D^{\alpha}
\Bigr)^2\right]\quad .
\end{equation}
The transverse projections of the moment taken at $\gamma=1$ define the 
complex {\it news function} of the electromagnetic waves\footnote{I am
using the terminology of Bondi [7].} 
\begin{equation}
\frac{\partial}{\partial u}\; m_{\alpha}\D^{\alpha}\biggl |_{\gamma=1}
\equiv \frac{\partial}{\partial u}\C(u,\phi)
\end{equation}
so that, by (1.20) and (1.27),
\begin{equation}
-\frac{\partial\e}{\partial u}=-\frac{1}{4\pi\kappa^2}\,\tr
\Bigl(\frac{\partial}{\partial u}\C\Bigr)
\Bigl(\frac{\partial}{\partial u}\C^*\Bigr)\quad.
\end{equation}
Finally, the longitudinal projection of the moment
\begin{equation}
\nabla_{\alpha}r\,\D^{\alpha}\equiv \D_{||}(u,\phi)
\end{equation}
plays no role in classical theory but, as shown below, it is responsible
for the vacuum creation of charge.

As pointed out in paper [2], the calculation in this paper is
insufficient for a removal of the limitation (1.24) and needs to be
revised. The revised calculation with all the needed amendments is
carried out in the present paper. Its result is that the quantity
(1.21) diverges {\it indeed}. The pole at $\gamma=1$ goes but its
place is taken up by an infrared divergence. Only the total energy
flux
\begin{equation}
\frac{\partial{\cal E}_{\mbox{\scriptsize tot}}}{\partial u}=
\frac{\partial\e}{\partial u}+\frac{\partial\E}{\partial u}
\end{equation}
is finite. The point here is that the electromagnetic field to be
inserted in (1.31) should solve the expectation-value equations.
To an appropriate order in the coupling constant,
$\partial\E/\partial u$ can be calculated with the bare source
but $\partial\e/\partial u$ should already be quantum corrected,
and this correction does not boil down to a renormalization of the
coupling constant. This correction is infrared divergent, and it
cancels the divergence in $\partial\E/\partial u$. The final result
is
\par\smallskip
\begin{eqnarray}
-\frac{\partial{\cal E}_{\mbox{\scriptsize tot}}}{\partial u}=
-\frac{1}{4\pi\kappa^2}\,\tr 
\Bigl(\frac{\partial}{\partial u}\C_{\mbox{\scriptsize eff}}\Bigr)
\Bigl(\frac{\partial}{\partial u}\C^*_{\mbox{\scriptsize eff}}\Bigr) 
\hspace{8cm}
\nonumber \\ 
{}-\frac{1}{(4\pi)^2}\frac{1}{3}\int\limits^1_0 d\gamma
\frac{\gamma^2}{(1-\gamma^2)}\,\tr\, g_{\alpha\beta}
\left[\Bigl(\frac{\partial}{\partial u}\D^{\alpha}\Bigr)
\Bigl(\frac{\partial}{\partial u}\D^{\beta}\Bigr)-
\Bigl(\frac{\partial}{\partial u}\D^{\alpha}\Bigr)
\Bigl(\frac{\partial}{\partial u}\D^{\beta}\Bigr)\biggl|_{\gamma=1}
\right] 
\end{eqnarray}
with the effective news function
\begin{eqnarray}
\frac{\partial}{\partial u}\C_{\mbox{\scriptsize eff}}(u,\phi)&=&
\Bigl[ 1-\frac{\kappa^2}{24\pi}({\bf c}-\log 2 +\frac{25}{12})\Bigr]
\frac{\partial}{\partial u}\C(u,\phi) \nonumber \\
&&{}-\frac{\kappa^2}{24\pi}\frac{\partial}{\partial u}\int
\limits^u_{-\infty} d\tau\log \bigl(m(u-\tau)\bigr)
\frac{\partial}{\partial \tau}\C(\tau,\phi)\quad . 
\end{eqnarray}
Here $\kappa^2$ is the renormalized coupling constant, ${\bf c}$ is the
Euler constant, and, as distinct from (1.21), the result is not independent
of the mass $m$ even in the high-frequency approximation. The quantities
$\D^{\alpha}$ and $\C$ in the expressions above pertain to the original
{\it bare} source.

The retarded integral along $\I$ in (1.33) represents a nonlocal tail
of the electromagnetic radiation. Technically, when a point tends to
$\I$, its past light cone becomes a sum of a null hyperplane and
a single null generator that merges with $\I$ [2]. The nonlocal
radiation tail is a contribution of this generator (see Appendix C).

Since the energy flux in (1.32) belongs partially to charged particles
and partially to the electromagnetic waves, the two contributions
should, of course, be separable:
\begin{equation}
\frac{\partial{\cal E}_{\mbox{\scriptsize tot}}}{\partial u}=
\frac{\partial{\cal E}_{\mbox{\scriptsize charge}}}{\partial u}+
\frac{\partial{\cal E}_{\mbox{\scriptsize waves}}}{\partial u}\quad .
\end{equation}
A calculation of the flux of charge helps to make this separation.
The density of the outgoing flux of charge can be calculated from
the expectation-value equations for the electromagnetic field
(Sec. 3 below). The result for this density reads
\begin{equation}
-\frac{\partial}{\partial u}{\hat e}(u,\phi)=
-\frac{\kappa^2}{3(4\pi)^2}\frac{\partial}{\partial u}\D_{||}
\biggl |_{\gamma=1}
\end{equation}
so that the total released charge is
\begin{equation}
\int\limits_{-\infty}^{\infty}du\sphere\left(-\frac{\partial}{\partial u}
{\hat e}\right) 
=\frac{\kappa^2}{3(4\pi)^2}\sphere \left[\D_{||}(u=-\infty)-
\D_{||}(u=+\infty)\right]\biggl |_{\gamma=1}\quad . 
\end{equation}
Hence one may infer that the portion of the total energy flux (1.32)
that goes with the charged particles is the one associated with
the longitudinal projection of the moment:
\begin{equation}
-\frac{\partial{\cal E}_{\mbox{\scriptsize charge}}}{\partial u}=
-\frac{1}{(4\pi)^2}\frac{1}{3}\int\limits^1_0 d\gamma\,\gamma^2\,
\tr\Bigl(\frac{\partial}{\partial u}\D_{||}\Bigr)^2\quad .
\end{equation}

The remaining energy in (1.32) goes with the electromagnetic radiation.
Since anyway expression (1.32) is valid only in the high-frequency
approximation, condition (1.1) can be used for its further simplification.
It will be recalled that the domain of nonstationarity of the bare
source is assumed compact. Its temporal scale (in time $u$) is a purely
classical quantity of order $1/\nu$. Therefore, if in (1.33) one writes
\begin{equation}
\log \bigl(m(u-\tau)\bigr)=\log\frac{m}{\nu} +\log\bigl(\nu(u-\tau)\bigr)
\quad ,
\end{equation}
the contribution of the second term will be of order $O(1)$ whereas
the contribution of the first term will be large:
\begin{equation}
\frac{\partial}{\partial u}\C_{\mbox{\scriptsize eff}}(u,\phi)=
\left(1-\frac{\kappa^2}{24\pi}\log\frac{m}{\nu}+
O\Bigl(\left(\frac{m}{\nu}\right)^0\Bigr)\right)\frac{\partial}{\partial u}
\C(u,\phi)\quad .
\end{equation}
As a result, for $u$ in the support of the bare news function, 
the radiation flux becomes merely a renormalized classical one:
\begin{equation}
\frac{\partial{\cal E}_{\mbox{\scriptsize waves}}}{\partial u}=
Z\frac{\partial\e}{\partial u}\quad ,
\end{equation}
\begin{equation}
Z=1-\frac{\kappa^2}{12\pi}\log\frac{m}{\nu}+
O\Bigl(\left(\frac{m}{\nu}\right)^0\Bigr)\quad .
\end{equation}
Note the sign of the quantum correction! The radiation gets amplified.

The results above pertain to the standard loop. For other models the
vacuum fluxes are multiples of the respective fluxes for the standard
loop (Sec. 8). Thus, for the spinor QED, the flux of charge is
{\it twice} the one in (1.35), and the quantum correction to the flux
of energy is {\it twice} the one in (1.32). 
Only the numerical constant which in
(1.33) is $25/12$ needs to be calculated anew but this constant is
anyway unimportant. The explicit results for the spinor QED are
obtained by introducing the said factor of 2 and substituting
\begin{equation}
\R_{\mu\nu}=-\mbox{i}qF_{\mu\nu}{\hat 1}\;\; ,\quad
{\hat e}=-\mbox{i}qe{\hat 1}\;\; , \quad
\kappa^2=4q^2\;\; ,\quad  \tr{\hat 1}=4 
\end{equation}
where $F_{\mu\nu}$ is the Maxwell tensor, $e$ is the electric charge
of the source, $q$ is the electron's charge, and $m$ in (1.33) is
the electron's mass.

In conclusion it will be noted that the result obtained cannot be
the end of the story since, obviously, it violates the energy
conservation law. Indeed, the frequency $\nu$ is proportional 
to the energy of the bare source, and, since the factor
$\log\nu/m$ can be however large, at a sufficiently large $\nu$
the source will radiate more energy than it has initially.
In this respect the present case is similar to the case of charged
spherical shell considered in Ref. [2]. A spherical shell expanding
in the self field emits no electromagnetic waves but, at an
ultrarelativistic energy, its vacuum radiation violates the energy
conservation law. The measure of the violation is in both cases
one and the same, $\kappa^2\log\nu$, and the cause is also one and 
the same: the problem has not been made fully self-consistent.
Although in the present case the backreaction of the vacuum on the
electromagnetic field is taken into account (otherwise the emitted
energy would not even be finite), its reaction on the motion of the
source is not. This task remains beyond the scope of the present
work but it may be conjectured that the missing backreaction
effect is nonanalytic in the coupling constant.

Eqs. (1.32)-(1.36) and their corollaries are the main results of 
the present work. Their derivation is given below. A reader not
interested in the technical details may still want to read
Secs. 2 to 4. Sec. 2 presents the general scheme of the calculation
including the important intermediate results and displays the mechanism
of the vacuum backreaction. Sec. 3 presents the solution of the
expectation-value equations and the calculation of the emission of charge.
In Sec. 4, creation of massive particles is considered, and a criterion
of the high-frequency approximation is derived.

The technical details are presented in Secs. 5 to 8. The calculation
required in the present work is more complicated than in [2] because
the nonlocal form factors act now on functions having noncompact
spatial supports. For a test function $X$, compactness of the spatial
support is equivalent to the following powers of decrease at null
infinities ${\cal I}^{\pm}$ and spatial infinity $\mbox{i}^0$:
\begin{equation}
X\biggl |_{{\cal I}^{\pm}}=O\Bigl(\frac{1}{r^3}\Bigr)\quad ,
\quad X\biggl |_{{\mbox i}^0}=O\Bigl(\frac{1}{r^4}\Bigr)\quad .
\end{equation}
The behaviours of the form factors derived or quoted in [2] are
valid only under conditions (1.43). For the present calculation, critical
is the behaviour of the test function at $\I$. The test function that
doesn't satisfy condition (1.43) at $\I$ will be called {\it singular}
at $\I$. The operators $\log(-\Box)$ and $1/\Box$ with test functions
singular at $\I$ are considered in Appendix C. The behaviours of the
third-order form factors at $\I$ are obtained in Appendix B. Appendix A
summarizes the structure of the one-loop form factors.

}

%% file: arttxt02.tex
{\renewcommand{\theequation}{2.\arabic{equation}}

\begin{center}
\section{\bf    The mechanism of the vacuum backreaction}
\end{center}

$$ $$

For obtaining the vacuum energy-momentum tensor to lowest order
in the commutator curvature, one needs the terms in the effective
action quadratic in the commutator curvature and linear in the
gravitational curvature, i.e. quadratic terms of order $\R\times\R$
and cubic terms of order $R\times\R\times\R$. Their general form is
[8,9]
\begin{equation}
\S=\S(2)+\S(3)+\mbox{higher-order terms}\quad ,
\end{equation}
\begin{eqnarray}
\S(2)&=&\frac{1}{2(4\pi)^2}\integral\tr\;\R_{\mu\nu}\gamma(-\Box)
\R^{\mu\nu}\quad ,\\
\S(3)&=&\frac{1}{2(4\pi)^2}\integral\tr\;\sum_i \Gamma_i 
(-\Box_1 ,-\Box_2 ,-\Box_3)R_1\R_2\R_3(i)
\end{eqnarray}
with some form factors $\gamma(-\Box)$ and
$\Gamma_i (-\Box_1 ,-\Box_2 ,-\Box_3)$. In the basis of nonlocal 
invariants of third order [9], there are 6 invariants of the needed
type:
\begin{eqnarray}
R_1\R_2\R_3(7)&=&R_1\R_2^{\;\mu\nu}\R_{3\mu\nu}\nonumber\\
R_1\R_2\R_3(8)&=&R_1^{\;\alpha\beta}\R_{2\alpha}{}^{\mu}\R_{3\beta\mu}
\nonumber\\
R_1\R_2\R_3(18)&=&R_{1\alpha\beta}\nabla_{\mu}\R_2^{\;\mu\alpha}
\nabla_{\nu}\R_3^{\;\nu\beta}\nonumber\\
R_1\R_2\R_3(19)&=&R_1^{\;\alpha\beta}\nabla_{\alpha}\R_2^{\;\mu\nu}
\nabla_{\beta}\R_{3\mu\nu}\nonumber\\
R_1\R_2\R_3(20)&=&R_1\nabla_{\alpha}\R_2^{\;\alpha\mu}\nabla^{\beta}
\R_{3\beta\mu}\nonumber\\
R_1\R_2\R_3(21)&=&R_1^{\;\mu\nu}\nabla_{\mu}\nabla_{\lambda}\R_2^{\;\lambda
\alpha}\R_{3\alpha\nu}
\end{eqnarray}
(I preserve the numbers that these invariants have in the full list
of Ref. [9]).

For $\kappa^2$ in (1.6) to be the renormalized coupling constant, the
form factor $\gamma(-\Box)$ should satisfy the normalization condition
$\gamma(0)=0$. The normalized $\gamma(-\Box)$ calculated for the
standard loop is
\begin{equation}
\gamma(-\Box)=-\frac{1}{2}\int\limits_0^1 d\alpha\int\limits_0^{
\alpha (1-\alpha)}dx\;\log\Bigl(1-x\frac{\Box}{m^2}\Bigr)\quad .
\end{equation}
When applied to a high-frequency field, this operator takes the form
\begin{equation}
\gamma(-\Box)=\frac{1}{12}\left(\frac{8}{3}-\log(-\frac{\Box}{m^2})\right)
+O(m^2)\quad .
\end{equation}
(The high-frequency limit is considered in Sec. 4.) The
constant $8/3$ in (2.6) is observable since it accounts for the 
difference between the static regime in which the total initially
stored charge is calculated and the high-frequency regime in which
the emission of charge is calculated. Neither this constant nor the
term in $\log m^2$ can be discarded when the operator (2.6) acts on
a function singular at $\I$ (cf. [2]).

The third-order form factors $\Gamma_i$ admit the massless limit and,
in the high-frequency approximation, can be taken massless from the
outset. One can then use the results of Ref. [9] where the massless
$\Gamma_i$ are calculated for all cubic invariants including the
ones in (2.4).

For obtaining $\T^{\mu\nu}$ at $\I$ one doesn't need the exact form 
factors. It suffices to have the asymptotic behaviours of
$\Gamma_i (-\Box_1 ,-\Box_2 ,-\Box_3)$ with one of the arguments
small and the others fixed. The difference with Ref. [2] is that
these behaviours are now needed including the terms $O(\Box^0)$.
The algorithms of extracting the needed terms are derived in
Appendix B.

The contribution of the second-order action (2.2) to $\T^{\mu\nu}$
will be divided into two:
\begin{equation}
\frac{2}{g^{1/2}}\frac{\delta\S(2)}{\delta g_{\mu\nu}}=
\T^{\mu\nu}(1)+\T^{\mu\nu}(2)
\end{equation}
with $\T^{\mu\nu}(2)$ the contribution of the variation of the
form factor:
\begin{equation}
\integral\T^{\mu\nu}(2)\;\delta g_{\mu\nu}=\frac{1}{(4\pi)^2}\integral
\tr\;\R_{\mu\nu}\,\delta\gamma(-\Box)\,\R^{\mu\nu}\quad .
\end{equation}
Denoting $\T^{\mu\nu}(3)$ the contribution of the third-order action
(2.3), one has
\begin{equation}
\T^{\mu\nu}=\T^{\mu\nu}(1)+\T^{\mu\nu}(2)+\T^{\mu\nu}(3)\quad .
\end{equation}
The vacuum energy flux in (1.10) will then also be a sum of the
respective three contributions:
\begin{equation}
\frac{\partial\E}{\partial u}=\frac{\partial\E(1)}{\partial u}+
\frac{\partial\E(2)}{\partial u}+\frac{\partial\E(3)}{\partial u}\quad .
\end{equation}

The expectation-value equations are obtained by varying the total
action $\s+\S$ with respect to the connection field. These are the
following equations for the source of the full, quantum-corrected,
commutator curvature:
\begin{equation}
\J^{\mu}_{\mbox{\scriptsize Full}}+\frac{\kappa^2}{2\pi}\gamma(-\Box)
\J^{\mu}_{\mbox{\scriptsize Full}}=
\J^{\mu}_{\mbox{\scriptsize Bare}}
\end{equation}
with the retarded boundary conditions for $\gamma(-\Box)$ [2].
Solving them iteratively one obtains
\begin{equation}
\J^{\mu}_{\mbox{\scriptsize Full}}=
\J^{\mu}_{\mbox{\scriptsize Bare}}-\frac{\kappa^2}{2\pi}
\gamma(-\Box)\J^{\mu}_{\mbox{\scriptsize Bare}}
\end{equation}
and hence
\begin{equation}
\R^{\mu\nu}_{\mbox{\scriptsize Full}}=
\R^{\mu\nu}_{\mbox{\scriptsize Bare}}-\frac{\kappa^2}{2\pi}
\gamma(-\Box)\R^{\mu\nu}_{\mbox{\scriptsize Bare}}\quad .
\end{equation}

For displaying the mechanism of the vacuum backreaction, it suffices
to write down the expressions for $\t^{\mu\nu}$ and $\T^{\mu\nu}(1)$ :
\begin{eqnarray}
\t^{\mu\nu}&=&-\frac{1}{4\pi\kappa^2}\,\tr\;\Bigl(\R^{\mu\lambda}
\R^{\nu}{}_{\lambda}-\frac{1}{4}g^{\mu\nu}\R_{\alpha\beta}
\R^{\alpha\beta}\Bigr)\quad ,\\
\T^{\mu\nu}(1)&=&-\frac{1}{8\pi^2}\,\tr\;\Bigl(\R^{\mu\lambda}\,\gamma(-\Box)\,
\R^{\nu}{}_{\lambda}-\frac{1}{4}g^{\mu\nu}\R_{\alpha\beta}
\,\gamma(-\Box)\,\R^{\alpha\beta}\Bigr)\quad .
\end{eqnarray}
Using Eq. (2.13) one finds
\begin{equation}
\t^{\mu\nu}\biggl |_{J=J_{\mbox{\scriptsize Full}}}=
\Bigl(\t^{\mu\nu}-2\T^{\mu\nu}(1)\Bigr)
\biggl |_{J=J_{\mbox{\scriptsize Bare}}}\quad .
\end{equation}
The total energy-momentum tensor of the commutator curvature,
$\t^{\mu\nu}+\T^{\mu\nu}$ , is then
\begin{eqnarray}
T^{\mu\nu}_{\mbox{\scriptsize tot}}&=&
\Bigl(\t^{\mu\nu}+\T^{\mu\nu}(1)+\T^{\mu\nu}(2)+\T^{\mu\nu}(3)\Bigr)
\biggl |_{J=J_{\mbox{\scriptsize Full}}}\nonumber \\
&=&\Bigl(\t^{\mu\nu}{\bf -}\T^{\mu\nu}(1)+\T^{\mu\nu}(2)+
\T^{\mu\nu}(3)\Bigr)\biggl |_{J=J_{\mbox{\scriptsize Bare}}}
\end{eqnarray}
and hence the total energy flux is
\begin{equation}
\frac{\partial {\cal E}_{\mbox{\scriptsize tot}}}{\partial u}=
\Bigl(\frac{\partial\e}{\partial u}{\bf -}\frac{\partial\E(1)}{\partial u}
+\frac{\partial\E(2)}{\partial u}+\frac{\partial\E(3)}{\partial u}\Bigr)
\biggl |_{J=J_{\mbox{\scriptsize Bare}}}\quad .
\end{equation}
Thus the effect of the vacuum backreaction is {\it changing the sign of}
$\T^{\mu\nu}(1)$. As will be seen in a moment, this effect is dramatic.

Note that if the substitution (2.13) was made in the action, then, after
varying with respect to $g_{\mu\nu}$, both $\T^{\mu\nu}(1)$ and
$\T^{\mu\nu}(2)$ would change their signs. This procedure is incorrect
because it amounts to varying the action in $g_{\mu\nu}$ at fixed
$\R_{\mbox{\scriptsize Bare}}$ whereas the energy-momentum tensor 
is obtained by varying the action in $g_{\mu\nu}$
at fixed $\R_{\mbox{\scriptsize Full}}$.
This makes difference since the relation between 
$\R_{\mbox{\scriptsize Full}}$ and $\R_{\mbox{\scriptsize Bare}}$ itself
depends on the metric through the operator $\Box$. The correct procedure
is making the substitution (2.13) in the energy-momentum tensor.

The dictum that $\T^{\mu\nu}$ at $\I$ is infrared divergent means that
expansion (1.10) does not hold. Rather there is an expansion of the
form
\begin{equation}
\T^{\mu\nu}\|=\mbox{terms}\;\frac{\log r}{r^2}+\mbox{terms}\;\frac{1}{r^2}
+O\Bigl(\frac{1}{r^3}\Bigr)\quad .
\end{equation}
If this was the behaviour of the total energy-momentum tensor, the
expectation-value spacetime would fail to be asymptotically flat. 
This is not the case but, in the intermediate expressions, the factor 
$\log r$ will conventionally be included in $\partial\E/\partial u$
thereby considering this energy flux as divergent.

The contributions (2.10) to $\partial\E/\partial u$ are calculated
in Secs. 5,6, and 7 below. Their main ingredient is the $\gamma=1$
radiation moment
\begin{equation}
\D^{\alpha}\biggl|_{\gamma=1}\equiv \d^{\alpha}\quad .
\end{equation}
The latter notation is used everywhere below. The results are as
follows:
\begin{eqnarray}
-\frac{\partial\E(1)}{\partial u}&=&\frac{1}{(4\pi)^2}\frac{1}{6}\,\tr
\Biggl\{ -(\log mr +2{\bf c}-\log 2 +\frac{8}{3})
\Bigl(\frac{\partial}{\partial u}\d^{\alpha}\Bigr)
\Bigl(\frac{\partial}{\partial u}\d{}_{\alpha}\Bigr)\nonumber \\
&&{}-\Bigl(\frac{\partial}{\partial u}\d^{\alpha}\Bigr)
\frac{\partial}{\partial u}\int\limits^u_{-\infty}d\tau\,\log
\bigl(m(u-\tau)\bigr)\frac{\partial}{\partial\tau}\d{}_{\alpha}(\tau)
\Biggr\}\quad , \\
-\frac{\partial\E(2)}{\partial u}&=&\frac{1}{(4\pi)^2}\frac{1}{6}\,\tr
\Biggl\{ \Bigl(\frac{\partial}{\partial u}\d^{\alpha}\Bigr)
\frac{\partial}{\partial u}\int\limits^u_{-\infty}d\tau\,\log
(u-\tau)\frac{\partial}{\partial\tau}\d{}_{\alpha}(\tau)\nonumber \\
&&{}-\frac{\partial}{\partial u}\int\limits^u_{-\infty}d\tau\,\log
(u-\tau)\Bigl(\frac{\partial}{\partial\tau}\d^{\alpha}(\tau)\Bigr)
\Bigl(\frac{\partial}{\partial\tau}\d{}_{\alpha}(\tau)\Bigr)\Biggr\}\nonumber \\
&&{}+\mbox{Q.N.}\quad ,\\
-\frac{\partial\E(3)}{\partial u}&=&\frac{1}{(4\pi)^2}\frac{1}{6}\,\tr
\Biggl\{ -(\log r +\log 2 -\frac{3}{2})
\Bigl(\frac{\partial}{\partial u}\d^{\alpha}\Bigr)
\Bigl(\frac{\partial}{\partial u}\d{}_{\alpha}\Bigr)\nonumber \\
&&{}+\frac{\partial}{\partial u}\int\limits^u_{-\infty}d\tau\,\log (u-\tau)
\Bigl(\frac{\partial}{\partial\tau}\d^{\alpha}(\tau)\Bigr)
\Bigl(\frac{\partial}{\partial\tau}\d{}_{\alpha}(\tau)\Bigr)\nonumber \\
&&{}-2\int\limits_0^1 d\gamma\,\frac{\gamma^2}{1-\gamma^2}
\left[\Bigl(\frac{\partial}{\partial u}\D^{\alpha}\Bigr)
\Bigl(\frac{\partial}{\partial u}\D_{\alpha}\Bigr)-
\Bigl(\frac{\partial}{\partial u}\d^{\alpha}\Bigr)
\Bigl(\frac{\partial}{\partial u}\d{}_{\alpha}\Bigr)\right]\Biggr\}\nonumber \\
&&{}+\mbox{Q.N.}\quad .
\end{eqnarray}
Owing to the conservation law (1.26), only the transverse projections of
$\d^{\alpha}$ survive in these expressions. Therefore, under the
limitation (1.24) the contributions $\partial\E(1)/\partial u$ and
$\partial\E(2)/\partial u$ vanish, and the contribution
$\partial\E(3)/\partial u$ gives back the result of Ref. [2].

When the limitation (1.24) does not hold, the contribution
$\partial\E(1)/\partial u$ is infrared divergent. The contribution
$\partial\E(2)/\partial u$ is not but it has another pathology.
The total-derivative term in (2.22) does not vanish in the integral 
over time. On the contrary, the behaviour of this term at late
time is 
\begin{eqnarray}
\frac{\partial}{\partial u}&
{\displaystyle \int\limits^u_{-\infty}d\tau\,\log (u-\tau)
\Bigl(\frac{\partial}{\partial\tau}\d^{\alpha}(\tau)\Bigr)
\Bigl(\frac{\partial}{\partial\tau}\d{}_{\alpha}(\tau)\Bigr)
\biggl|_{u\to\infty}} \nonumber \\
&{\displaystyle =\frac{1}{u}\int\limits^{\infty}_{-\infty}d\tau\,
\Bigl(\frac{\partial}{\partial\tau}\d^{\alpha}(\tau)\Bigr)
\Bigl(\frac{\partial}{\partial\tau}\d{}_{\alpha}(\tau)\Bigr)}
\end{eqnarray}
so that the integrated flux (2.22) diverges:
\begin{equation}
\int\limits^{\infty}_{-\infty}du\,\frac{\partial\E(2)}{\partial u}
=\infty\quad .
\end{equation}
The contribution $\partial\E(3)/\partial u$ contains the divergences
of both types. In the sum of the three contributions the divergence
of the integral in time cancels but the infrared divergence 
{\it doubles}:
\begin{equation}
-\frac{\partial\E}{\partial u}=-\frac{1}{(4\pi)^2}\frac{1}{3}
(\log r)\;\tr\,\Bigl(\frac{\partial}{\partial u}\d^{\alpha}\Bigr)
\Bigl(\frac{\partial}{\partial u}\d{}_{\alpha}\Bigr)+O(1)\quad .
\end{equation}
Only in the total sum (2.18) with the changed sign of 
$\partial\E(1)/\partial u$ both divergences cancel, and the finite
result (1.32) emerges.

The cancellations outlined above do not depend on the relative sign
and coefficient between $\s$ and $\S$ (the $\kappa^2$ in (1.6) is
in fact kept arbitrary) but they depend crucially on the balance
between $\S(2)$ and $\S(3)$. As seen from the expressions
(2.21)-(2.23), there is a precise relation between the respective
contributions, and this relation maintains for other field models
(Sec. 8) despite the fact that $\S(2)$ emerges from the purely
electromagnetic coupling whereas $\S(3)$ represents the vertices
with the gravitational coupling. Owing to this relation, the final
result is rigidly tied to the overall coefficient of the action
$\S(2)$ which is merely the $\beta$-function. A knowledge of
this coefficient is in the end sufficient for obtaining the
vacuum radiation fluxes.

}

%% file: arttxt03.tex
{\renewcommand{\theequation}{3.\arabic{equation}}

\begin{center}
\section{\bf    The mean electromagnetic field. Emission\protect\\of charge}
\end{center}

$$ $$

Since the quantum correction to the electromagnetic field cancels
the infrared divergence in the vacuum energy, it should itself be infrared
divergent. This point is clarified below but, before considering
the expectation-value equations, it is useful to make a general
analysis of the asymptotic properties of the commutator curvature
and its source in the case where there is an emission of both
waves and charge. To make difference with the notation already used,
the quantities in this analysis will be distinguished with boldface.

The existence of a flux of charge at a large distance from the source 
implies that ${\bf\J^{\alpha}}$ falls off at $\I$ like
\begin{equation}
{\bf\J^{\alpha}}\|=\frac{{\bf j^{\alpha}}(u,\phi)}{r^2}+O\Bigl(
\frac{1}{r^3}\Bigr)
\end{equation}
with some coefficient ${\bf j^{\alpha}}(u,\phi)$. This is the most general
behaviour admissible for an isolated system. Although the support of the
source ${\bf\J^{\alpha}}$ is no more confined to a spacetime tube,
its domain of nonstationarity must remain compact in time in order
that all fluxes die out in the past and future of $\I$. More generally,
the source should be asymptotically stationary in the past and future.
To account for this property in the past, 
it will be assumed that the domain of nonstationarity
of ${\bf\J^{\alpha}}$ is confined to the interior of some future
light cone $u=u_-$. Then
\begin{equation}
{\bf j^{\alpha}}(u,\phi)\biggl|_{u<u_-}=0\quad .
\end{equation}

The density of the flux of charge from a source is expressed through
the coefficient in (3.1) as follows [2]:
\begin{equation}
-\frac{\partial}{\partial u}{\bf{\hat e}}(u,\phi)=\frac{1}{8\pi}
\nabla_{\alpha}v\;(r^2{\bf\J^{\alpha}})\|=\frac{1}{8\pi}\nabla_{\alpha}v\;
{\bf j^{\alpha}}(u,\phi)
\end{equation}
with $\nabla v$ in (1.11). The function
\begin{equation}
{\bf{\hat e}}(u)\equiv {\bf{\hat e}}(-\infty)+\int\limits^u_{-\infty}
d{\bar u}\sphere\frac{\partial}{\partial{\bar u}}{\bf{\hat e}}
({\bar u},\phi)
\end{equation}
defined by (3.3) up to an additive constant ${\bf{\hat e}}(-\infty)$
can be written as an integral over the future light cone [2]:
\begin{equation}
{\bf{\hat e}}(u)=\frac{1}{4\pi}\int d{\bar x}\,{\bar g}^{1/2}
\delta ({\bar u}-u){\bar\nabla}_{\bar\mu}{\bar u}\;{\bf\J^{\bar\mu}}
({\bar x})
\end{equation}
provided that the constant
\begin{equation}
{\bf{\hat e}}\equiv {\bf{\hat e}}(-\infty)
\end{equation}
is taken as a conserved integral over an arbitrary (complete) spacelike
hypersurface:
\begin{equation}
{\bf{\hat e}}=\frac{1}{4\pi}\int d{\bar x}\,{\bar g}^{1/2}
\delta \bigl(\tau({\bar x})\bigr){\bar\nabla}_{\bar\mu}\tau({\bar x})\;
{\bf\J^{\bar\mu}}({\bar x})\;\; ,\quad (\nabla\tau)^2<0\quad .
\end{equation}
The function ${\bf{\hat e}}(u)$ may be called the Bondi charge, and
the constant ${\bf{\hat e}}$ the ADM charge since their meaning is
the same as of the Bondi and ADM masses\footnote{I continue using the
terminology of the theory of asymptotically flat spaces [4].}. The
ADM charge is the total charge of the source ${\bf\J}$ in the initial
state, i.e. before the beginning of emission. The Bondi charge is the
charge that remains in a compact domain by the instant $u$ of retarded
time in the process of emission. The ADM charge is conserved because
at any instant $u$ it equals a sum of the charge emitted by this
instant and the charge remaining by this instant, Eq. (3.4).

Consider the conservation equation
\begin{equation}
\nabla_{\alpha}{\bf\J^{\alpha}}=0\quad .
\end{equation}
Inserting the expansion (3.1) in (3.8) one obtains
\begin{equation}
\frac{\partial}{\partial u}\;(\nabla_{\alpha}u\;{\bf j^{\alpha}})=0
\end{equation}
whence, in view of (3.2),
\begin{equation}
\nabla_{\alpha}u\;{\bf j^{\alpha}}=0\quad .
\end{equation}
The latter equation makes it possible to express the flux of charge
in (3.3) through the longitudinal projection of ${\bf j^{\alpha}}$:
\begin{equation}
-\frac{\partial}{\partial u}{\bf{\hat e}}(u,\phi)=
\frac{1}{4\pi}\nabla_{\alpha}r\;{\bf j^{\alpha}}(u,\phi)\quad .
\end{equation}

Thus the longitudinal projection
$\nabla_{\alpha}r\:{\bf j^{\alpha}}$ of the residue in (3.1) is
responsible for the emission of charge, the projection
$\nabla_{\alpha}u\:{\bf j^{\alpha}}$ vanishes but no conclusion
can be made on the transverse projections $m_{\alpha}\,{\bf j^{\alpha}}$.
Their vanishing does not follow and, at this stage, their role
remains unclear.

For obtaining the behaviour of the commutator curvature one must
first consider the question of convergence of the moment
${\bf\d^{\alpha}}$ of the source ${\bf\J}$. The analysis of convergence is 
carried out in [2]. When applied to the present case, it gives
the following result. The projection of ${\bf\d^{\alpha}}$ on a
basis vector in (1.25) converges if and only if the {\it like}
projection of the residue ${\bf j^{\alpha}}$ vanishes. It follows that the 
projection $\nabla_{\alpha}u\:{\bf\d^{\alpha}}$ converges, the
projection $\nabla_{\alpha}v\:{\bf\d^{\alpha}}$ diverges, and
the behaviours of the transverse projections 
$m_{\alpha}{\bf\d^{\alpha}}$ remain undetermined. Hence using the
results for the retarded operator $1/\Box$ in Ref. [2] and Appendix C
below one obtains
\begin{eqnarray}
\nabla_{\alpha}u\;\frac{1}{\Box}{\bf\J^{\alpha}}\|&=&-\frac{1}{r}
\nabla_{\alpha}u\;{\bf\d^{\alpha}}(u,\phi)\quad ,\\
\nabla_{\alpha}v\;\frac{1}{\Box}{\bf\J^{\alpha}}\|&=&-\frac{\log r}{r}
\frac{1}{2}\int\limits^u_{-\infty}d{\bar u}\;\nabla_{\alpha}v\,
{\bf j^{\alpha}}({\bar u},\phi)
\end{eqnarray}
while for the transverse projections one has two cases:
\begin{eqnarray}
{\rm i})\;\; m_{\alpha}\,{\bf j^{\alpha}}=0\;\; ,\qquad\qquad m_{\alpha}
\frac{1}{\Box}{\bf\J^{\alpha}}\| &=&-\frac{1}{r} m_{\alpha}
{\bf\d^{\alpha}}(u,\phi)\quad ,\qquad \\
{\rm ii})\;\; m_{\alpha}\,{\bf j^{\alpha}}\ne 0\;\; ,\qquad\qquad m_{\alpha}
\frac{1}{\Box}{\bf\J^{\alpha}}\| &=&-\frac{\log r}{r}\frac{1}{2}
\int\limits^u_{-\infty}d{\bar u}\; m_{\alpha}\,{\bf j^{\alpha}}
({\bar u},\phi)\quad .\qquad
\end{eqnarray}

When the support of the source is confined to a spacetime tube,
the convergent projection $\nabla_{\alpha}u\:{\bf\d^{\alpha}}$
equals the total charge of the source (Eq. (1.26)). This projection
remains the conserved ADM charge also in the general case, even
when there is an emission of charge and despite the fact that
the integration hypersurface in ${\bf\d^{\alpha}}$ is null:
\begin{equation}
\frac{\partial}{\partial u}\left(\nabla_{\alpha}u\;
{\bf\d^{\alpha}}\right)=0\quad ,\quad \nabla_{\alpha}u\;
{\bf\d^{\alpha}}={\bf{\hat e}}\quad.
\end{equation}
The proof uses the explicit form of the null hyperplane [2]
and the stationarity of the source in the past.

The fact that the projection $\nabla_{\alpha}v\:{\bf\d^{\alpha}}$
is generally divergent presents no real problem since this projection 
drops out of both the square of the differentiated moment in (1.27)
and the commutator curvature. Indeed, solving the Jacobi identities
to lowest order [2], one obtains for the commutator curvature
\begin{equation}
{\bf{\hat R}}_{\mu\nu}=\nabla_{\nu}\frac{1}{\Box}{\bf\J_{\mu}}-
\nabla_{\mu}\frac{1}{\Box}{\bf\J_{\nu}}
\end{equation}
whence
\begin{equation}
{\bf{\hat R}}_{\mu\nu}\|=\frac{\partial}{\partial u}\Bigl(
\nabla_{\nu}u\,\frac{1}{\Box}{\bf\J_{\mu}}-\nabla_{\mu}u\,
\frac{1}{\Box}{\bf\J_{\nu}}\Bigr)+O\Bigl(\frac{1}{r^2}\Bigr)\quad .
\end{equation}
The projection (3.13) drops out of this expression by symmetry.
Moreover, owing to the conservation law (3.16) one finds
\begin{eqnarray}
{\bf{\hat R}}_{\mu\nu}\|&=&\frac{1}{2}\frac{\partial}{\partial u}
\left [(\nabla_{\nu}u\;m_{\mu}-\nabla_{\mu}u\;m_{\nu})
\Bigl(m^*_{\alpha}\frac{1}{\Box}{\bf\J^{\alpha}}\Bigr)\right. \\
&&\quad\; {}+\left.(\nabla_{\nu}u\;m^*_{\mu}-\nabla_{\mu}u\;m^*_{\nu})
\Bigl(m_{\alpha}\frac{1}{\Box}{\bf\J^{\alpha}}\Bigr)\right]+
O\Bigl(\frac{1}{r^2}\Bigr)\quad,\nonumber
\end{eqnarray}
and the only projection of ${\bf{\hat R}}_{\mu\nu}$ that can behave like
$1/r$ (counting only powers) is
\begin{equation}
\frac{1}{2}\nabla^{\mu}v\;m^{\nu}\;{\bf{\hat R}}_{\mu\nu}\|=
\frac{\partial}{\partial u}\Bigl(m_{\alpha}\frac{1}{\Box}
{\bf\J^{\alpha}}\Bigr)\|
\equiv {}-\frac{1}{r}\frac{\partial}{\partial u}{\bf{\hat C}}+
O\Bigl(\frac{1}{r^2}\Bigr)
\end{equation}
where the coefficient $\partial{\bf{\hat C}}/\partial u$ at $1/r$ will
conventionally be called news function although in the case ii) above
it is infrared divergent. One has either
\begin{equation}
{\rm i})\;\; m_{\alpha}\,{\bf j^{\alpha}}=0\;\; ,\qquad\qquad
\frac{\partial}{\partial u}
{\bf{\hat C}}=\frac{\partial}{\partial u}\Bigl(m_{\alpha}
{\bf\d^{\alpha}}(u,\phi)\Bigr) \qquad\quad\;\qquad\qquad\qquad
\end{equation}
or
\begin{equation}
{\rm ii})\;\; m_{\alpha}\,{\bf j^{\alpha}}\ne 0\;\; ,\qquad\qquad
\frac{\partial}{\partial u}
{\bf{\hat C}}=(\log r)\frac{1}{2}m_{\alpha}\,{\bf j^{\alpha}}(u,\phi)\quad .
\qquad\qquad\qquad\qquad
\end{equation}

At this stage there appears an argument to make a conclusion on the
transverse fluxes $m_{\alpha}\,{\bf j^{\alpha}}$. If one wants the news 
function to be finite, these fluxes must vanish. However, the only
reason for insisting that the news function be finite is making
finite the {\it energy} of the electromagnetic field since by (2.14)
and (3.20)
\begin{equation}
-\frac{\partial\e}{\partial u}=-\frac{1}{16\pi\kappa^2}\,\tr
\Bigl(r\;\nabla_{\mu}v\;m_{\nu}\;{\bf{\hat R}}^{\mu\nu}\Bigr)
\Bigl(r\;\nabla_{\alpha}v\;m^*_{\beta}\;{\bf{\hat R}}^{\alpha\beta}
\Bigr)\| 
\equiv{}-\frac{1}{4\pi\kappa^2}\,\tr
\Bigl(\frac{\partial}{\partial u}{\bf{\hat C}}\Bigr)
\Bigl(\frac{\partial}{\partial u}{\bf{\hat C}}^*\Bigr)\quad .
\end{equation}
This is the reason indeed but only if the energy-momentum tensor
of the electromagnetic field is given by expression (2.14). For
a classical field it is. Therefore, for a classical field one has
the case i) $m_{\alpha}\,{\bf j^{\alpha}}=0$, i.e. the only nonvanishing
flux of ${\bf\J^{\alpha}}$ is the flux of charge in (3.11), and the only
divergent projection of the moment is 
$\nabla_{\alpha}v\:{\bf\d^{\alpha}}$. The flux of the electromagnetic
energy is then completely determined by the finite news function in (3.21).
However, if a c-number electromagnetic field is an expectation value 
rather than the classical field, its energy-momentum tensor is
{\it not} (2.14). Rather it is a sum
$\t^{\mu\nu}+\T^{\mu\nu}$~, and the same argument that the energy
should be finite may now be in favour of the case ii)
$m_{\alpha}\,{\bf j^{\alpha}}\ne 0$ where the news function is divergent.

One is now ready to consider the expectation-value equations. In the
high-frequency approximation, Eq. (2.12) takes the form
\begin{equation}
\J^{\mu}_{\mbox{\scriptsize Full}}=\J^{\mu}_{\mbox{\scriptsize Bare}}
-\frac{\kappa^2}{24\pi}\left(\frac{8}{3}-\log(-\frac{\Box}{m^2})\right)
\J^{\mu}_{\mbox{\scriptsize Bare}}\quad .
\end{equation}
Since the bare source has a compact
spatial support, one can use the result from Ref. [2]:
\begin{equation}
\log(-\Box)X\|=-\frac{2}{r^2}\frac{\partial}{\partial u}
D_{\bf 1}(u,\phi|X)+O\Bigl(\frac{1}{r^3}\Bigr)
\end{equation}
which is valid under conditions (1.43) and in which
$D_{\bf 1}(u,\phi|X)$ is the $\gamma=1$ moment of the test source $X$.
Since the local terms in (3.24) are $O(1/r^3)$, one obtains
\begin{equation}
\J^{\alpha}_{\mbox{\scriptsize Full}}\|=-\frac{1}{r^2}
\frac{\kappa^2}{12\pi}\frac{\partial}{\partial u}\d^{\alpha}{}_{\;
\mbox{\scriptsize Bare}}+O\Bigl(\frac{1}{r^3}\Bigr)\quad .
\end{equation}
This is Eq. (3.1) with
\begin{equation}
{\bf j^{\alpha}}(u,\phi)=-\frac{\kappa^2}{12\pi}
\frac{\partial}{\partial u}\d^{\alpha}{}_{\;\mbox{\scriptsize Bare}}\quad .
\end{equation}
Hence, using the conservation law (3.16) for the bare source, one obtains
\begin{equation}
\nabla_{\alpha}u\;{\bf j^{\alpha}}=-\frac{\kappa^2}{12\pi}
\frac{\partial}{\partial u}\Bigl(\nabla_{\alpha}u\;\d^{\alpha}{}_{\;
\mbox{\scriptsize Bare}}\Bigr)=0
\end{equation}
and thereby checks condition (3.10). Next, using Eq. (3.11) one calculates
the density of the flux of charge
\begin{equation}
-\frac{\partial}{\partial u}{\hat e}_{\;\mbox{\scriptsize Full}}(u,\phi)=
-\frac{\kappa^2}{3(4\pi)^2}\frac{\partial}{\partial u}
\Bigl(\nabla_{\alpha}r\;\d^{\alpha}{}_{\;\mbox{\scriptsize Bare}}\Bigr)
\end{equation}
and thereby obtains the result (1.35). Finally, one calculates
the transverse fluxes
\begin{equation}
m_{\alpha}\,{\bf j^{\alpha}}=-\frac{\kappa^2}{12\pi}
\frac{\partial}{\partial u}\Bigl(m_{\alpha}\d^{\alpha}{}_{\;
\mbox{\scriptsize Bare}}\Bigr)=-\frac{\kappa^2}{12\pi}
\frac{\partial}{\partial u}\C_{\mbox{\scriptsize Bare}}
\end{equation}
and discovers that they are proportional to the news function of the
bare source. It follows that if the bare source emits waves, then
the news function of the full source diverges. By (3.22),
\begin{equation}
\frac{\partial}{\partial u}\C_{\mbox{\scriptsize Full}}=
\left(1-(\log r)\frac{\kappa^2}{24\pi}\right)
\frac{\partial}{\partial u}\C_{\mbox{\scriptsize Bare}}+O(1)\quad .
\end{equation}
However, one knows already that this divergence comes to the rescue.
One can now check this again. From (3.23) and (3.31) one obtains
\begin{equation}
\frac{\partial\e}{\partial u}\biggl|_{J_{\mbox{\scriptsize Full}}}=
\frac{1}{4\pi\kappa^2}\Bigl(1-\frac{\kappa^2}{12\pi}\log r +
\kappa^2 O(1)\Bigr)\;\tr
\Bigl(\frac{\partial}{\partial u}\C_{\mbox{\scriptsize Bare}}\Bigr)
\Bigl(\frac{\partial}{\partial u}\C^*_{\mbox{\scriptsize Bare}}\Bigr)\quad .
\end{equation}
On the other hand, by (2.26),
\begin{equation}
\frac{\partial\E}{\partial u}=\frac{1}{3(4\pi)^2}(\log r)\;\tr
\Bigl(\frac{\partial}{\partial u}\C_{\mbox{\scriptsize Bare}}\Bigr)
\Bigl(\frac{\partial}{\partial u}\C^*_{\mbox{\scriptsize Bare}}\Bigr)+
O(1)\quad .
\end{equation}
As a result, the total energy flux (1.31) is finite.

The approximate form (3.24) of the expectation-value equations 
corresponds to a neglect of the mass of the vacuum particles and is 
valid only in the region $u>u_-$ where the source is assumed
strongly variable. This form can be used for a calculation of the
derivative of the Bondi charge in the high-frequency approximation
but cannot be used for a calculation of the ADM charge since the
latter calculation involves the region $u<u_-$ where the source
is static. The {\it electro}static polarization with massless
vacuum particles is infinite. Indeed, for the ADM charge (3.7)
to converge, the full source must fall off at {\it spatial} 
infinity like
\begin{equation}
{\bf\J}\;\biggl|_{\mbox{i}^0}=O\Bigl(\frac{1}{r^4}\Bigr)
\end{equation}
whereas a calculation with the massless form factor $\log(-\Box)$
in (3.24) yields the behaviour
\begin{equation}
\log(-\Box)\,\J_{\mbox{\scriptsize Bare}}\;\biggl|_{\mbox{i}^0}=
O\Bigl(\frac{1}{r^3}\Bigr)
\end{equation}
and the divergent result\footnote{The only exception is the case where the
bare source has no monopole moment, ${\hat e}_{\;\mbox{\scriptsize Bare}}
=0$. Then one can show that also ${\hat e}_{\;\mbox{\scriptsize Full}}
=0$. An {\it observable} electric charge cannot be carried by massless
particles.}
\begin{equation}
{\hat e}_{\;\mbox{\scriptsize Full}}=\left(1-(\log r)
\frac{\kappa^2}{12\pi}\right){\hat e}_{\;\mbox{\scriptsize Bare}}+O(1)\;\; ,
\quad m=0\;\; .
\end{equation}
The correct result for the ADM charge is obtained with the normalized
massive form factor (2.5):
\begin{equation}
{\hat e}_{\;\mbox{\scriptsize Full}}={\hat e}_{\;\mbox{\scriptsize Bare}}
\;\; ,\qquad m\ne 0\;\; .
\end{equation}

}

%% file: arttxt04.tex
{\renewcommand{\theequation}{4.\arabic{equation}}

\begin{center}
\section{\bf    Creation of massive particles and the 
high-frequency approximation}
\end{center}

$$ $$

In spite of their apparent similarity, the divergent renormalization
(3.36) of the ADM charge and the divergent renormalization (3.31)
of the news function have different status. The former is a result of 
an incorrect use of the high-frequency approximation in the static
region whereas the latter is a natural consequence of the intensive
pair creation. To show this and to derive a criterion of the
high-frequency approximation, the expectation-value equations are
considered below with the massive form factor $\gamma(-\Box)$.

The kernel of the operator (2.5) is obtained with the aid of 
its spectral form
\begin{equation}
\gamma(-\Box)=\frac{1}{12}\Biggl[\frac{8}{3}+\int\limits^{\infty}_{4m^2}
d\mu^2\,
\frac{\Bigl(1-\frac{\textstyle 4m^2}{\textstyle\mu^2}\Bigr)^{3/2}}{
\mu^2-\Box}-
\int\limits_0^{\infty}d\mu^2\,\frac{1}{\mu^2+m^2}\Biggr]
\end{equation}
in which the threshold $\mu^2=4m^2$ appears explicitly, and, for the
convergence at the upper limit, the two spectral integrals are to be
considered as a single integral. Outside the support of the source
$\J_{\mbox{\scriptsize Bare}}$ , Eq. (2.12) takes the form
\begin{equation}
\J^{\alpha}_{\mbox{\scriptsize Full}}=-\frac{\kappa^2}{24\pi}
\int\limits_{4m^2}^{\infty}d\mu^2\,\Bigl(1-\frac{4m^2}{\mu^2}\Bigr)^{3/2}
\frac{1}{\mu^2-\Box}\J^{\alpha}_{\mbox{\scriptsize Bare}}
\end{equation}
with the retarded resolvent $(\mu^2-\Box)^{-1}$.
Here the order of integrations is important [10]. The spacetime
integration implied in 
$(\mu^2-\Box)^{-1}\J_{\mbox{\scriptsize Bare}}$ is to be done first,
and the spectral-mass integration next.

One is presently interested in the behaviour of the full source 
at a large distance from the support tube of 
$\J_{\mbox{\scriptsize Bare}}$ . At $r\gg l$ and $u$ fixed, the
retarded resolvent acting on a nonstationary source behaves as follows [10]:
\begin{equation}
\frac{1}{\mu^2-\Box}\J^{\alpha}_{\mbox{\scriptsize Bare}}\:\propto\:
\frac{1}{r}\exp \left(-\mu\sqrt{f(u)\, r}\right)
\end{equation}
where $f(u)$ is a positive function of time and angles having
the dimensions $1/\nu$.

Using (4.2) and (4.3) one can estimate the fluxes associated with
created particles. For $r\gg l$ one finds
\begin{eqnarray}
\J^{\alpha}_{\mbox{\scriptsize Full}}\:&\propto&\:
\frac{1}{r}\int\limits_{4m^2}^{\infty}d\mu^2\,
\Bigl(1-\frac{4m^2}{\mu^2}\Bigr)^{3/2}
\exp\left(-\mu\sqrt{f(u)\,r}\right)\nonumber \\
&=&\frac{1}{r^2}\frac{1}{f(u)}\int\limits_0^{\infty}dx^2\,
\frac{x^3}{(x^2+4m^2\,f(u)\,r)^{3/2}}\exp\left(
-\sqrt{x^2+4m^2\,f(u)\,r}\right)\quad .
\end{eqnarray}
When projected on $\nabla_{\alpha}r$, the coefficient of $1/r^2$ in
the latter expression is the density of the flux of charge through 
a tube of radius $r$. It follows that, because of the presence of 
the threshold, the flux through the tube of radius $r\gg l$ is
suppressed by the factor
\begin{equation}
\exp\Bigl(-\frac{2m\sqrt{r}}{\sqrt{\nu}}\Bigr)\quad .
\end{equation}

Hence one infers that, although pair creation starts as soon as
$\hbar\nu$ reaches the value of order $mc^2$, the particles are
created in the support of the source with small momenta and
don't get far away. They stay in a compact spatial domain until
$\hbar\nu$ reaches the value
\begin{equation}
\hbar\nu\sim mc^2\left(\frac{mc}{\hbar}l\right)\quad .
\end{equation}
At this value there appears an observable flux of charged particles
outside the support tube of $\J_{\mbox{\scriptsize Bare}}$ . The
factor $(mc/\hbar)l$ may be interpreted as the number of created
particles for which there is room in the spatial support of the
source. If the creation is more violent, the particles get out
of the tube. Finally, the high-frequency approximation is valid
when $\hbar\nu$ is much bigger than (4.6):
\begin{equation}
\hbar\nu\gg mc^2\left(\frac{mc}{\hbar}l\right)\quad .
\end{equation}
Under this condition the flux of created energy and charge stops
depending on the mass of the particles. The mass terms in (4.4)
can then be discarded which is equivalent to replacing the second
term of the spectral formula (4.1) with
\begin{equation}
\int\limits_0^{\infty}d\mu^2\,\frac{1}{\mu^2-\Box}\quad .
\end{equation}
Hence the approximation (2.6) for the form factor.

Expression (4.4) holds for the transverse projections of 
$\J_{\mbox{\scriptsize Full}}$ as well:
\begin{equation}
m_{\alpha}\J^{\alpha}_{\mbox{\scriptsize Full}}\;
\biggl |_{\textstyle r\gg l}
\;\propto\frac{1}{r^2}\quad ,
\end{equation}
and the coefficient of $1/r^2$ in this expression is nonvanishing
whenever there is a nonvanishing flux of created particles
outside the support of $\J_{\mbox{\scriptsize Bare}}$ . Only in
the special case where the electromagnetic radiation of
$\J_{\mbox{\scriptsize Bare}}$ is absent altogether may the
transverse projections vanish. Thus the behaviour (4.9) is a
direct consequence of pair creation. Then, by (3.22), the news
function of the mean field inevitably diverges like $\log r$.
Hence relation (3.31). The normalization scale of the $\log r$
in this relation can be read from the kernel of the operator
$1/\Box$ (Appendix C). This is $\log (r/l)$.

It will be emphasized once again that there is nothing wrong
about the mean electromagnetic field. When pairs are created,
its energy is no more governed by its news function since there
appears a real vacuum contribution. The news function diverges but 
the vacuum energy redistributes
and keeps the electromagnetic radiation down.

}

%% file: arttxt05.tex
{\renewcommand{\theequation}{5.\arabic{equation}}

\begin{center}
\section{\bf    Calculation of \protect\(\protect\T\protect^{
\protect\mu\protect\nu}(1)\protect\) 
at \protect\(\protect\I\protect\)}
\end{center}

$$ $$

There remains to be presented the calculation of the energy fluxes
(2.21), (2.22), and (2.23). This is, of course, the main part of
the work.

The $\T^{\mu\nu}(1)$ is given by expression (2.15) with 
$\gamma(-\Box)$ in (2.6). The commutator curvature to be inserted in
$\T^{\mu\nu}$ is the one generated by the bare source. The bare
source has a compact spatial support. Therefore [2],
\begin{equation}
-\frac{1}{\Box}\J^{\alpha}_{\mbox{\scriptsize Bare}}\;\|=
\frac{1}{r}\d^{\alpha}+O\Bigl(\frac{1}{r^2}\Bigr)
\end{equation}
and hence by (3.18)
\begin{equation}
\R^{\mu\nu}\|=\frac{1}{r}\Bigl(\nabla^{\mu}u\;\frac{\partial}{\partial u}
\d^{\nu}-\nabla^{\nu}u\;\frac{\partial}{\partial u}\d^{\mu}\Bigr)+
O\Bigl(\frac{1}{r^2}\Bigr)\quad .
\end{equation}

It follows that for obtaining $\T^{\mu\nu}(1)$ at $\I$ one needs to know
the behaviour of $\log(-\Box)$ with a test function that behaves at $\I$
like
\begin{equation}
X\|=\frac{1}{r}A(u,\phi)+O\Bigl(\frac{1}{r^2}\Bigr)\quad .
\end{equation}
The needed result is obtained in Appendix C:
\begin{eqnarray}
-\log\Bigl(-\frac{\Box}{m^2}\Bigr)X\|&=&\frac{A(u,\phi)}{r}
(\log mr +2{\bf c}-\log 2 )\nonumber \\
&&{}+\frac{1}{r}\int\limits^u_{-\infty}d\tau\,\log\bigl(m(u-\tau)\bigr)
\frac{\partial}{\partial\tau}A(\tau,\phi)+O\Bigl(\frac{\log r}{r^2}
\Bigr)\quad .
\end{eqnarray}

Substituting (5.2) for (5.3) one obtains
\begin{eqnarray}
-\R^{\mu}{}_{\lambda}\:\log\Bigl(-\frac{\Box}{m^2}\Bigr)\,\R^{\nu\lambda}
\|=\frac{1}{r^2}\nabla^{\mu}u\nabla^{\nu}u\;\Biggl [
(\log mr+2{\bf c}-\log 2)\Bigl(\frac{\partial}{\partial u}\d^{\alpha}
\Bigr)\Bigl(\frac{\partial}{\partial u}\d{}_{\alpha}\Bigr)\nonumber \\
{}+\Bigl(\frac{\partial}{\partial u}\d^{\alpha}\Bigr)
\int\limits^u_{-\infty}d\tau\,\log\bigl(m(u-\tau)\bigr)
\frac{\partial^2}{\partial\tau^2}\d{}_{\alpha}(\tau)\Biggr ]
+O\Bigl(\frac{1}{r^3}\Bigr)\quad .
\end{eqnarray}
Here and below, use is to be made of the following identity:
\begin{equation}
\frac{d}{du}\int\limits^u_{-\infty}d\tau\,\log(u-\tau)\,f(\tau)=
\int\limits^u_{-\infty}d\tau\,\log(u-\tau)
\frac{d}{d\tau}f(\tau)
\end{equation}
where $f(\tau)$ is supposed to provide the convergence at the 
lower limit. The convergence of the integral in (5.5) and similar
integrals is provided by the assumption of asymptotic stationarity
of the bare source. Under the simplified assumption that the domain
of nonstationarity of the source is compact, there will be time
instants $u_-$ and $u_+$ such that~[2]
\begin{equation}
\frac{\partial}{\partial u}\D^{\alpha}(u)\biggl |_{\textstyle u<u_-}=0
\quad ,\quad \frac{\partial}{\partial u}\D^{\alpha}(u)
\biggl |_{\textstyle u>u_+}=0\quad .
\end{equation}

In this way the result (2.21) is obtained.

}

%% file: arttxt06.tex
{\renewcommand{\theequation}{6.\arabic{equation}}

\begin{center}
\section{\bf    Calculation of \protect\(\protect\T\protect^{
\protect\mu\protect\nu}(2)\protect\) 
at \protect\(\protect\I\protect\)}
\end{center}

$$ $$

The $\T^{\mu\nu}(2)$ is defined by Eq. (2.8) with $\gamma(-\Box)$
in (2.6). Using the spectral formula
\begin{equation}
\log\Bigl(-\frac{\Box}{m^2}\Bigr)=-\int\limits_0^{\infty}d\mu^2\,
\biggl(\frac{1}{\mu^2-\Box}-\frac{1}{\mu^2+m^2}\biggr)
\end{equation}
one finds
\begin{equation}
\integral\T^{\mu\nu}(2)\delta g_{\mu\nu}=\frac{1}{12(4\pi)^2}
\integral\,\tr\int\limits_0^{\infty}d\mu^2\,
\Bigl(\frac{1}{\mu^2-\Box}\R^{\alpha\beta}\Bigr)\delta\Box
\Bigl(\frac{1}{\mu^2-\Box}\R_{\alpha\beta}\Bigr)\quad ,
\end{equation}
and the operator $\delta\Box$ can be obtained by calculating
\begin{equation}
\integral\,\tr\;\R^{\alpha\beta}(\delta\Box)\R_{\alpha\beta}=
\integral\,\delta g_{\mu\nu}\;\tr\biggl({\LARGE p}^{\mu\nu}
(\nabla_1 ,\nabla_2 )\R_1^{\;\alpha\beta}\R_{2\alpha\beta}\biggr)
\end{equation}
where ${\LARGE p}^{\mu\nu}(\nabla_1 ,\nabla_2 )$ is some polynomial
in the derivative $\nabla_1$ acting on $\R_1$ and the derivative
$\nabla_2$ acting on $\R_2$. In terms of this operator polynomial
one obtains
\begin{equation}
\T^{\mu\nu}(2)=-\frac{1}{12(4\pi)^2}\,\tr\;{\LARGE p}^{\mu\nu}
(\nabla_1 ,\nabla_2 )\frac{\log(\Box_1/\Box_2)}{\Box_1-\Box_2}
\R_1^{\;\alpha\beta}\R_{2\alpha\beta}
\end{equation}
where the nonlocal form factor results from the spectral-mass
integration in (6.2), and, up to higher orders in the curvature,
the operators in (6.4) are commutative.

The explicit form of Eq. (6.3) is
\begin{eqnarray}
\integral\,\tr\;\R^{\alpha\beta}(\delta\Box)\R_{\alpha\beta}=
\integral\,\delta g_{\mu\nu}\;\tr\biggl[
\Bigl(\nabla^{\mu}\R^{\alpha\beta}\Bigr)
\Bigl(\nabla^{\nu}\R_{\alpha\beta}\Bigr)-\frac{1}{4} g^{\mu\nu}\Box
\Bigl(\R^{\alpha\beta}\R_{\alpha\beta}\Bigr)\nonumber \\
-2\nabla_{\alpha}\Bigl(\R^{\mu}{}_{\beta}\nabla^{\nu}\R^{\alpha\beta}
-\R^{\alpha\beta}\nabla^{\nu}\R^{\mu}{}_{\beta}\Bigr)\biggr]\hspace{15mm}
\end{eqnarray}
whence
\begin{equation}
\T^{\mu\nu}(2)=-\frac{1}{12(4\pi)^2}\,\tr\,\frac{\log(\Box_1/\Box_2)}{
\Box_1-\Box_2}\nabla_1^{\;\mu}\R_1^{\;\alpha\beta}\;\nabla_2^{\;\nu}
\R_{2\alpha\beta}+\mbox{total derivatives}\quad .
\end{equation}
A detailed analysis shows that the total derivatives in (6.6) either
vanish at $\I$ or vanish in the integrated energy flux, i.e. their
contribution to $\partial\E/\partial u$ is quantum noise (Eq. (1.22)).
The technique used in this analysis is outlined below.

In the remaining term of (6.6) use will be made of Eq. (3.17) and the
conservation equation (3.8) to express the commutator curvature through 
its source:
\begin{equation}
\R_1^{\;\alpha\beta}\R_{2\alpha\beta}=\Box\Bigl(\frac{1}{\Box_1\Box_2}
\J_1^{\;\alpha}\J_{2\alpha}\Bigr)-2\nabla_{\alpha}\nabla_{\beta}
\Bigl(\frac{1}{\Box_1\Box_2}\J_1^{\;\alpha}\J_2^{\;\beta}\Bigr)
-\frac{1}{\Box_1}\J_1^{\;\alpha}\J_{2\alpha}-
\J_1^{\;\alpha}\frac{1}{\Box_2}\J_{2\alpha}\quad .
\end{equation}
The first two terms of this expression are total derivatives. One obtains
\begin{equation}
\T^{\mu\nu}(2)=\frac{1}{6(4\pi)^2}\,\tr\;\frac{1}{\Box_2}\;
\frac{\log(\Box_1/\Box_2)}{\Box_1-\Box_2}\nabla_1^{\;(\mu}\J_{1\alpha}
\nabla_2^{\;\nu)}\J_2^{\;\alpha}+\mbox{total derivatives}
\end{equation}
where the indices $\mu\nu$ are symmetrized. Of the new total derivatives,
the contribution of the first term in (6.7) is $O(1/r^3)$ at $\I$,
and the contribution of the second term is quantum noise. The proof is
given below.

The form factor in (6.8) can be expressed through the operator
${\cal H}_q$ introduced in [2] and Appendix A below:
\begin{equation}
\frac{1}{\Box_2}\;\frac{\log(\Box_1/\Box_2)}{\Box_1-\Box_2}X_1X_2(x)=
-\int\limits^0_{-\infty}dq\,\biggl(\frac{d}{dq}{\cal H}_qX_1(x)\biggr)
\biggl(\int\limits^q_{-\infty}\frac{d{\bar q}}{{\bar q}}
{\cal H}_{{\bar q}}X_2(x)\biggr)\quad .
\end{equation}
The behaviour of this function as $x\to\I$ is obtained in the same way
as in [2] by making the replacement of the integration variable
\begin{equation}
q=r(\tau-u)\quad ,\quad r=r(x)\to\infty
\end{equation}
where $\tau$ is the new integration variable, and $u=u(x)$ is the retarded
time of the point $x$ at $\I$. With $q$ replaced as in (6.10), one has [2]
\begin{equation}
{\cal H}_qX(x)\biggl|_{x\to\I}=\frac{1}{r}D_{\bf 1}(\tau ,\phi|X)
\end{equation}
where the quantity on the right-hand side is the $D_{\bf 1}$ moment
of the test source X. As a result, for (6.9) one obtains
\begin{equation}
\frac{1}{\Box_2}\;\frac{\log(\Box_1/\Box_2)}{\Box_1-\Box_2}X_1X_2\|=
-\frac{1}{r^2}\int\limits^u_{-\infty}d\tau\,
\biggl(\frac{\partial}{\partial\tau}D_{\bf 1}(\tau ,\phi|X_1)\biggr)
\biggl(\int\limits^{\tau}_{-\infty}\frac{d{\bar\tau}}{{\bar\tau}-u}
D_{\bf 1}({\bar\tau},\phi|X_2)\biggr)\quad .
\end{equation}

Using the latter result in (6.8) one finds
\begin{eqnarray}
\T^{\mu\nu}(2)\|=\frac{1}{r^2}\nabla^{\mu}u\nabla^{\nu}u\;
\frac{1}{6(4\pi)^2}\,\tr\,\Biggl[-\int\limits^u_{-\infty}d\tau\,
\biggl(\frac{\partial^2}{\partial\tau^2}\d^{\alpha}(\tau)\biggr)
\biggl(\int\limits^{\tau}_{-\infty}\frac{d{\bar\tau}}{{\bar\tau}-u}
\frac{\partial}{\partial{\bar\tau}}\d{}_{\alpha}({\bar\tau})\biggr)
\nonumber \\
{}+\mbox{Q.N.}\Biggl]+O\Bigl(\frac{1}{r^3}\Bigr)\quad .
\end{eqnarray}
The integration by parts first in the internal integral
\begin{equation}
\int\limits^{\tau}_{-\infty}\frac{d{\bar\tau}}{{\bar\tau}-u}
\frac{\partial}{\partial{\bar\tau}}\d{}_{\alpha}({\bar\tau})=
\log (u-\tau)\frac{\partial}{\partial\tau}\d{}_{\alpha}(\tau)
-\int\limits^{\tau}_{-\infty}d{\bar\tau}\,\log(u-{\bar\tau})
\frac{\partial^2}{\partial{\bar\tau}^2}\d{}_{\alpha}({\bar\tau})
\end{equation}
and next in the external integral yields finally the expression
(2.22).

It is now convenient to present a proof that the first two terms
in (6.7) can be discarded. Their contributions to $\T^{\mu\nu}(2)$
are respectively
\begin{equation}
\Delta_1\T^{\mu\nu}=-\frac{1}{12(4\pi)^2}\,\tr\;\Box
\biggl(\frac{1}{\Box_1\Box_2}\;\frac{\log(\Box_1/\Box_2)}{\Box_1-\Box_2}
\nabla_1^{\;\mu}\J_1^{\;\alpha}\nabla_2^{\;\nu}\J_{2\alpha}\biggr)
\end{equation}
and
\begin{equation}
\Delta_2\T^{\mu\nu}=\frac{1}{6(4\pi)^2}\,\tr\;\nabla_{\alpha}\nabla_{\beta}
\biggl(\frac{1}{\Box_1\Box_2}\;\frac{\log(\Box_1/\Box_2)}{\Box_1-\Box_2}
\nabla_1^{\;\mu}\J_1^{\;\alpha}\nabla_2^{\;\nu}\J_2^{\;\beta}\biggr)\quad .
\end{equation}
Using the same technique as above, one obtains
\begin{eqnarray}
\frac{1}{\Box_1\Box_2}\;\frac{\log(\Box_1/\Box_2)}{\Box_1-\Box_2}
\nabla_1^{\;\mu}\J_1^{\;\alpha}\nabla_2^{\;\nu}\J_2^{\;\beta}\|\hspace{8cm}
\nonumber \\
=-\frac{1}{2r}\nabla^{\mu}u\nabla^{\nu}u\int\limits^u_{-\infty}d\tau\,
\biggl(\int\limits^{\tau}_{-\infty}\frac{d{\bar\tau}}{{\bar\tau}-u}
\frac{\partial}{\partial{\bar\tau}}\d^{\alpha}({\bar\tau})\biggr)
\biggl(\int\limits^{\tau}_{-\infty}\frac{d{\bar\tau}}{{\bar\tau}-u}
\frac{\partial}{\partial{\bar\tau}}\d^{\beta}({\bar\tau})\biggr)
+O\Bigl(\frac{1}{r^2}\Bigr)\; .
\end{eqnarray}
Note that this behaviour is not even $1/r^2$. It is $1/r$. Nevertheless,
one has
\begin{equation}
\Box\; O\Bigl(\frac{1}{r}\Bigr)\|=O\Bigl(\frac{1}{r^3}\Bigr)
\end{equation}
(see Appendix B). Therefore, the contribution (6.15) is indeed
$O(1/r^3)$ at $\I$.

To calculate the contribution (6.16) at $\I$, one may use the 
following result from [2]. If a symmetric tensor $V^{\alpha\beta}$
is analytic at $\I$, and its projections tangential to $\I$ vanish:
\begin{equation}
\nabla_{\alpha}u\;V^{\alpha\beta}\|=O\Bigl(\frac{1}{r}V\Bigr)\quad ,
\quad \nabla_{\alpha}u\nabla_{\beta}u\;V^{\alpha\beta}\|=
O\Bigl(\frac{1}{r^2}V\Bigr)\quad ,
\end{equation}
then
\begin{equation}
\nabla_{\alpha}\nabla_{\beta}V^{\alpha\beta}\|=\frac{1}{r}
\frac{\partial}{\partial u}\Bigl(g_{\alpha\beta}V^{\alpha\beta}\Bigr)
+O\Bigl(\frac{1}{r^2}V\Bigr)\quad .
\end{equation}
By (3.16), the tensor (6.17) possesses the properties of $V^{\alpha\beta}$.
As a result, for the contribution (6.16) one finds
\begin{equation}
\Delta_2\T^{\mu\nu}\|=\frac{1}{r^2}\nabla^{\mu}u\nabla^{\nu}u\;
\Bigl(-\Delta_2\frac{\partial\E}{\partial u}\Bigr)+
O\Bigl(\frac{1}{r^3}\Bigr)
\end{equation}
where the respective energy flux is a total derivative in time:
\begin{equation}
-\Delta_2\frac{\partial\E}{\partial u}=\frac{-1}{12(4\pi)^2}\,\tr\;
\frac{\partial}{\partial u}\int\limits^u_{-\infty}d\tau\,
\biggl(\int\limits^{\tau}_{-\infty}\frac{d{\bar\tau}}{{\bar\tau}-u}
\frac{\partial}{\partial{\bar\tau}}\d^{\alpha}({\bar\tau})\biggr)
\biggl(\int\limits^{\tau}_{-\infty}\frac{d{\bar\tau}}{{\bar\tau}-u}
\frac{\partial}{\partial{\bar\tau}}\d{}_{\alpha}({\bar\tau})\biggr)\quad .
\end{equation}
Since
\begin{eqnarray}
\int\limits^u_{-\infty}d\tau\,
\biggl(\int\limits^{\tau}_{-\infty}\frac{d{\bar\tau}}{{\bar\tau}-u}
\frac{\partial}{\partial{\bar\tau}}\d^{\alpha}({\bar\tau})\biggr)
\biggl(\int\limits^{\tau}_{-\infty}\frac{d{\bar\tau}}{{\bar\tau}-u}
\frac{\partial}{\partial{\bar\tau}}\d{}_{\alpha}({\bar\tau})\biggr)
\biggl|_{u\to\infty}\hspace{3cm}\nonumber \\
=\frac{1}{u}\Bigl(\d(+\infty)-\d(-\infty)\Bigr)^2\to 0\quad ,
\end{eqnarray}
one obtains
\begin{equation}
\int\limits^{\infty}_{-\infty}du\,\Bigl(\Delta_2
\frac{\partial\E}{\partial u}\Bigr)=0\quad .
\end{equation}
Thus (6.21) is quantum noise indeed.

}

%% file: arttxt07.tex
{\renewcommand{\theequation}{7.\arabic{equation}}

\begin{center}
\section{\bf    Calculation of \protect\(\protect\T\protect^{
\protect\mu\protect\nu}(3)\protect\) 
at \protect\(\protect\I\protect\)}
\end{center}

$$ $$

The $\T^{\mu\nu}(3)$ is obtained by varying the third-order
action (2.3) with respect to the metric. Only the Ricci curvature
that enters the basis invariants (2.4) needs to be varied. The
commutator curvatures in these invariants are to be expressed
through their sources via Eqs. (3.17) and (3.8). The result may be 
represented in the form
\begin{equation}
\T^{\mu\nu}(3)=\frac{1}{(4\pi)^2}\,\tr\;\sum_l {\tilde\Gamma}_l
(-\Box ,-\Box_2 ,-\Box_3 )\:\R_2\R_3{}^{\mu\nu}(l)
\end{equation}
where the form factors
${\tilde\Gamma}_l (-\Box ,-\Box_2 ,-\Box_3 )$ are linear combinations
of $\Gamma_i (-\Box ,-\Box_2 ,-\Box_3 )$, and the structures
$\R_2\R_3{}^{\mu\nu}(l)$ make some nonlocal tensor basis second-order
in (the source of) the commutator curvature. The first argument
$\Box$ of the form factors ${\tilde\Gamma}_l$ is the operator
argument that in the action $\S(3)$ acts on the Ricci curvature.
In the variational derivative (7.1) it becomes an {\it overall}
operator acting at the observation point. In the diagrammatic
language, the argument $\Box$ corresponds to the external line
of the current (7.1). This is explained in more detail in Appendix B.

It is shown in Appendix B that only the small-$\Box$ expansion of
${\tilde\Gamma}_l(-\Box ,-\Box_2 ,-\Box_3 )$ in the argument $\Box$
is relevant to the behaviour of the current (7.1) at $\I$. This
expansion has the form
\begin{equation}
{\tilde\Gamma}_l (-\Box ,-\Box_2 ,-\Box_3 )=\log (-\Box){\cal A}_l
(\Box_2 ,\Box_3 )+{\cal B}_l (\Box_2 ,\Box_3 )+O(\Box)\;\; ,\quad
\Box\to 0
\end{equation}
and the contribution at $\I$ of the terms $O(\Box)$ is already
$O(1/r^3)$. By the results in Appendix B, both 
${\cal A}_l (\Box_2 ,\Box_3 )$ and ${\cal B}_l (\Box_2 ,\Box_3 )$
can be expressed through the operator ${\cal H}_q$ in a way
similar to Eq. (6.9). Specifically, all
${\cal A}_l (\Box_2 ,\Box_3)$ are linear combinations of the
following operators $F_{mn}(\Box_2 ,\Box_3 )$:
\begin{equation}
F_{mn}(\Box_2 ,\Box_3)X_2X_3=-2\int\limits^0_{-\infty}dq\,
q^{m+n}\Bigl[\Bigl(\frac{d}{dq}\Bigr)^{m+1}{\cal H}_qX_2\Bigr]
\Bigl[\Bigl(\frac{d}{dq}\Bigr)^{n+1}{\cal H}_qX_3\Bigr]\quad .
\end{equation}
Moreover, the term with $\log(-\Box)$ in (7.2) always appears in the
combination
\begin{equation}
\log(-\Box)F_{mn}(\Box_2 ,\Box_3 )+L_{mn}(\Box_2 ,\Box_3 )
\end{equation}
with
\begin{equation}
L_{mn}(\Box_2 ,\Box_3 )X_2X_3=-2\int\limits^0_{-\infty}dq\,
\left(\log(-\frac{q}{2})+2{\bf c}\right)q^{m+n}
\Bigl[\Bigl(\frac{d}{dq}\Bigr)^{m+1}{\cal H}_qX_2\Bigr]
\Bigl[\Bigl(\frac{d}{dq}\Bigr)^{n+1}{\cal H}_qX_3\Bigr]\; .
\end{equation}
All ${\cal B}_l (\Box_2 ,\Box_3 )$ in (7.2) are linear combinations of
$L_{mn}(\Box_2 ,\Box_3 )$ and $F_{mn}(\Box_2 ,\Box_3 )$.

Thus one arrives at the ansatz
\begin{equation}
\T^{\mu\nu}(3)\|=\frac{1}{(4\pi)^2}\,\tr\;\sum_l\biggl(\log(-\Box)
{\cal A}_l (\Box_2 ,\Box_3 )+{\cal B}_l (\Box_2 ,\Box_3 )\biggr)
\R_2\R_3{}^{\mu\nu}(l)+O\Bigl(\frac{1}{r^3}\Bigr)
\end{equation}
in which both the basis elements and the operator coefficients
are presently to be determined. One can check that the following
9 structures quadratic in the source of the commutator curvature 
make a basis:
\begin{eqnarray}
\R_2\R_3{}^{\mu\nu}(1)&=&\J_2^{\;\mu}\cdot\J_3^{\;\nu} \\
\R_2\R_3{}^{\mu\nu}(2)&=&\frac{1}{2}\nabla^{\mu}\nabla^{\nu}
(\J_2^{\;\alpha}\cdot\J_{3\alpha}) \\
\R_2\R_3{}^{\mu\nu}(3)&=&\J_2^{\;\alpha}\cdot\nabla_3^{\;\mu}
\nabla_3^{\;\nu}\J_{3\alpha} \\
\R_2\R_3{}^{\mu\nu}(4)&=&\frac{1}{2}\nabla^{\mu}\nabla^{\nu}
\nabla_{\alpha}\nabla_{\beta}(\J_2^{\;\alpha}\cdot\J_3^{\;\beta}) \\
\R_2\R_3{}^{\mu\nu}(5)&=&\nabla_{\alpha}\nabla_{\beta}
(\J_2^{\;\alpha}\cdot\nabla_3^{\;\mu}\nabla_3^{\;\nu}\J_3^{\;\beta}) \\
\R_2\R_3{}^{\mu\nu}(6)&=&\nabla^{\alpha}(\nabla_2^{\;(\mu}\J_{2\alpha}
\cdot\J_3^{\;\nu)}) \\
\R_2\R_3{}^{\mu\nu}(7)&=&\nabla^{\alpha}(\J_{2\alpha}\cdot
\nabla_3^{\;(\mu}\J_3^{\;\nu)}) \\
\R_2\R_3{}^{\mu\nu}(8)&=&g^{\mu\nu}\J_2^{\;\alpha}\cdot\J_{3\alpha} \\
\R_2\R_3{}^{\mu\nu}(9)&=&g^{\mu\nu}\nabla_{\alpha}\nabla_{\beta}
(\J_2^{\;\alpha}\cdot\J_3^{\;\beta})\quad .
\end{eqnarray}
The last two structures will be omitted since they cannot contribute
to the energy flux through $\I$.

The respective coefficients
${\cal A}_l (\Box_2 ,\Box_3 )$ and ${\cal B}_l (\Box_2 ,\Box_3 )$ are
obtained using the algorithms of Appendix B and the table of the
third-order form factors in [9]. Only the basis element (7.10) with
$l=4$ has a nonvanishing ${\cal A}_l (\Box_2 ,\Box_3 )$:
\begin{eqnarray}
{\cal A}_l (\Box_2 ,\Box_3 )&=&0\;\; ,\quad l\ne 4 \\
{\cal A}_4 (\Box_2 ,\Box_3 )&=&-\frac{1}{3}\;\frac{1}{\Box_2\Box_3}
\Bigl(F_{22}(\Box_2 ,\Box_3 )-2F_{11}(\Box_2 ,\Box_3 )\Bigr)\quad .
\end{eqnarray}
This agrees with Ref. [2] where only the term with $\log(-\Box)$ in
(7.6) was considered. The results for ${\cal B}_l (\Box_2 ,\Box_3 )$
are as follows:
\begin{eqnarray}
{\cal B}_1 (\Box_2 ,\Box_3 )&=&0\quad , \\
{\cal B}_2 (\Box_2 ,\Box_3 )&=&-\frac{1}{6}\;\frac{1}{\Box_2\Box_3}+
\frac{1}{12}\biggl(\frac{1}{\Box_2}+\frac{1}{\Box_3}\biggr)
F_{11}(\Box_2 ,\Box_3 )\quad , \\
{\cal B}_3 (\Box_2 ,\Box_3 )&=&\frac{1}{12}\biggl(\frac{1}{\Box_2}-
\frac{1}{\Box_3}\biggr)\Bigl(F_{11}(\Box_2 ,\Box_3 )-
F_{00}(\Box_2 ,\Box_3 )\Bigr)\quad , 
\end{eqnarray}
\begin{equation}
{\cal B}_4 (\Box_2 ,\Box_3 )={\cal B'}_4 (\Box_2 ,\Box_3 )+
{\cal B''}_4 (\Box_2 ,\Box_3 )\quad ,
\end{equation}
\begin{eqnarray}
{\cal B'}_4 (\Box_2 ,\Box_3 )&=&-\frac{1}{3}\;\frac{1}{\Box_2\Box_3}
\Bigl(L_{22}(\Box_2 ,\Box_3 )-2L_{11}(\Box_2 ,\Box_3 )\Bigr)\quad , \\
{\cal B''}_4 (\Box_2 ,\Box_3 )&=&\frac{2}{3}\;\frac{1}{\Box_2\Box_3}\;
F_{11}(\Box_2 ,\Box_3 )\quad , \\
{\cal B}_5 (\Box_2 ,\Box_3 )&=&\frac{1}{6}\;\frac{1}{\Box_2\Box_3}
\Bigl(F_{12}(\Box_2 ,\Box_3 )-F_{21}(\Box_2 ,\Box_3 )\nonumber \\
&&\hspace{2cm}{}-F_{11}(\Box_2 ,\Box_3 )-F_{00}(\Box_2 ,\Box_3 )\Bigr)\quad ,\\
{\cal B}_6 (\Box_2 ,\Box_3 )&=&-\frac{1}{6}\biggl(\frac{1}{\Box_2}-
3\:\frac{1}{\Box_3}\biggr)F_{11}(\Box_2 ,\Box_3)-
\frac{1}{6}\;\frac{1}{\Box_2}\;F_{00}(\Box_2 ,\Box_3 )\quad , \\
{\cal B}_7 (\Box_2 ,\Box_3 )&=&\frac{1}{3}\;\frac{1}{\Box_3}
\Bigl(F_{11}(\Box_2 ,\Box_3 )-\frac{1}{2}F_{00}(\Box_2 ,\Box_3 )\Bigr)\quad .
\end{eqnarray}
Only ${\cal B}_4$ contains the contribution of $L_{mn}(\Box_2 ,\Box_3 )$
because the latter can appear only in the combination (7.4).

The kernels for the superpositions of the operators 
$F_{mn}(\Box_2 ,\Box_3 )$ with $1/\Box_2$ and $1/\Box_3$ are given
in Eqs. (A.15) and (A.16) of Appendix A. By the same calculation 
as in Eqs. (6.9)-(6.12) their behaviours at $\I$ are expressed
through the moments $D_{\bf 1}$ of the test functions:
\begin{eqnarray}
\frac{1}{\Box_3}\;F_{mn}(\Box_2 ,\Box_3 )X_2X_3\|\hspace{10cm}\nonumber \\
=-\frac{1}{r^2}\int\limits^u_{-\infty}d\tau\,(\tau-u)^{m+n}
\biggl(\frac{\partial^{m+1}}{\partial\tau^{m+1}}
D_{\bf 1}(\tau,\phi|X_2)\biggr)
\biggl(\frac{\partial^n}{\partial\tau^n}\int\limits^{\tau}_{-\infty}
\frac{d{\bar\tau}}{{\bar\tau}-u}D_{\bf 1}({\bar\tau},\phi|X_3)\biggr)\; ,
\end{eqnarray}
\begin{eqnarray}
\frac{1}{\Box_2\Box_3}\;F_{mn}(\Box_2 ,\Box_3 )X_2X_3\|\hspace{10cm}\nonumber \\
=-\frac{1}{2r}\int\limits^u_{-\infty}d\tau\,(\tau-u)^{m+n}
\biggl(\frac{\partial^m}{\partial\tau^m}\int\limits^{\tau}_{-\infty}
\frac{d{\bar\tau}}{{\bar\tau}-u}D_{\bf 1}({\bar\tau},\phi|X_2)\biggr)
\biggl(\frac{\partial^n}{\partial\tau^n}\int\limits^{\tau}_{-\infty}
\frac{d{\bar\tau}}{{\bar\tau}-u}D_{\bf 1}({\bar\tau},\phi|X_3)\biggr)\quad
\end{eqnarray}
\begin{equation}
\frac{1}{\Box_2\Box_3}\;X_2X_3\|=\frac{1}{r^2}D_{\bf 1}(u,\phi|X_2)
D_{\bf 1}(u,\phi|X_3)\quad.
\end{equation}
With these behaviours, the calculation of the ${\cal B}_l$ terms in
(7.6) essentially repeats the calculation in Sec. 6. The ${\cal B}_l$
terms that involve the form factors (7.27) are analogous to (6.8),
and the ${\cal B}_l$ terms that involve the form factors (7.28) are
analogous to (6.16). This concerns all ${\cal B}_l$ except ${\cal B'}_4$.

For $l\ne 4$ the results are as follows. Since ${\cal B}_1=0$, there
remain two basis structures, $l=6$ and $l=7$, in which the indices of 
the energy-momentum tensor do not belong to derivatives. Their
contributions at $\I$ vanish by virtue of the conservation law (3.16).
This may be exemplified with just one term
\begin{eqnarray}
\frac{1}{\Box_3}\;F_{11}(\Box_2 ,\Box_3 )\J_{2\alpha}\nabla_3^{\;\alpha}
\J_3^{\;\nu}\|&=&-\frac{1}{r^2}\nabla^{\alpha}u\int\limits^u_{-\infty}
d\tau\,(\tau-u)\Bigl(\frac{\partial^2}{\partial\tau^2}\d{}_{\alpha}
(\tau)\Bigr)\Bigl(\frac{\partial}{\partial\tau}\d^{\nu}(\tau)\Bigr)\nonumber \\
&&{}+O\Bigl(\frac{1}{r^3}\Bigr)=O\Bigl(\frac{1}{r^3}\Bigr)\quad .
\end{eqnarray}
The contributions of the remaining structures are of the form
\begin{equation}
{\cal B}_l (\Box_2 ,\Box_3 )\:\R_2\R_3{}^{\mu\nu}(l)\|=
\frac{1}{r^2}\nabla^{\mu}u\nabla^{\nu}u\;\biggl[-c\;\biggl|_l
\Bigl(\frac{\partial}{\partial u}\d^{\alpha}\Bigr)
\Bigl(\frac{\partial}{\partial u}\d{}_{\alpha}\Bigr)+
\mbox{Q.N.}\biggr]+O\Bigl(\frac{1}{r^3}\Bigr)
\end{equation}
with
\begin{equation}
c\;\Bigl|_{l=2}=\frac{1}{12}\quad ,\quad c\;\Bigl|_{l=3}=\frac{1}{6}
\quad ,\quad c\;\Bigl|_{l=5}=\frac{1}{4}\quad .
\end{equation}
The contribution of ${\cal B''}_4$ is also of the form (7.31) with
$c\bigl|_{l=4}=0$. {\it Thus the effect of all structures induced
by the third-order action except the basis structure with} $l=4$
{\it boils down to a finite renormalization of the classical news
function.}

The main contribution comes from the basis structure (7.10) with
$l=4$. One may write
\begin{equation}
\biggl(\log(-\Box){\cal A}_4 (\Box_2 ,\Box_3 )+{\cal B'}_4 (\Box_2 ,\Box_3 )
\biggr)\R_2\R_3{}^{\mu\nu}(4)=\nabla^{\mu}\nabla^{\nu}
\Bigl(\log(-\Box){\hat I}(x)+{\hat N}(x)\Bigr)
\end{equation}
where
\begin{eqnarray}
{\hat I}(x)&=&-\frac{1}{6}\nabla_{\alpha}\nabla_{\beta}\;\frac{1}{\Box_2\Box_3}
\Bigl(F_{22}(\Box_2 ,\Box_3 )-2F_{11}(\Box_2 ,\Box_3 )\Bigr)
\J_2^{\;\alpha}\J_3^{\;\beta}\quad , \\
{\hat N}(x)&=&-\frac{1}{6}\nabla_{\alpha}\nabla_{\beta}\;\frac{1}{\Box_2\Box_3}
\Bigl(L_{22}(\Box_2 ,\Box_3 )-2L_{11}(\Box_2 ,\Box_3 )\Bigr)
\J_2^{\;\alpha}\J_3^{\;\beta}\quad .
\end{eqnarray}
The scalar $\tr\;{\hat I}(x)$ is the central object in Ref. [2].
In the same way as above one obtains
\begin{eqnarray}
{\hat I}(x)\|&=&\frac{1}{6r^2}\int\limits^u_{-\infty}d\tau\,(u-\tau)
\Bigl(\frac{\partial}{\partial\tau}\d^{\alpha}(\tau)\Bigr)
\Bigl(\frac{\partial}{\partial\tau}\d{}_{\alpha}(\tau)\Bigr)
+O\Bigl(\frac{1}{r^3}\Bigr)\quad , \\
{\hat N}(x)\|&=&\frac{\log r}{6r^2}\int\limits^u_{-\infty}d\tau\,(u-\tau)
\Bigl(\frac{\partial}{\partial\tau}\d^{\alpha}(\tau)\Bigr)
\Bigl(\frac{\partial}{\partial\tau}\d{}_{\alpha}(\tau)\Bigr)\nonumber \\
&&{}+\frac{1}{6r^2}\int\limits^u_{-\infty}d\tau\,(u-\tau)
\left(\log (u-\tau)-\log 2 +2{\bf c}+\frac{1}{2}\right)
\Bigl(\frac{\partial}{\partial\tau}\d^{\alpha}(\tau)\Bigr)
\Bigl(\frac{\partial}{\partial\tau}\d{}_{\alpha}(\tau)\Bigr)\nonumber \\
&&{}+\frac{1}{12r^2}\d^{\alpha}(u)\d{}_{\alpha}(u)+
O\Bigl(\frac{1}{r^3}\Bigr)
\end{eqnarray}
where (7.36) reproduces the result in [2]. However, the scalar
$\tr\;{\hat I}(x)$ is needed at one more limit which in [2] is called 
${\rm i}^+$. This is the limit $r\to\infty$ along the timelike
geodesic that, when traced to the future, reaches infinity at the point
$\phi$ of the celestial sphere with the energy 
$E=(1-\gamma^2)^{-1/2}$ per unit rest mass. The result obtained in [2]
for this limit is
\begin{equation}
{\hat I}(x)\biggl|_{\mbox{i}^+}=\frac{\gamma(1-\gamma^2)}{12r}
\int\limits^{\infty}_{-\infty}du\,
\Bigl(\frac{\partial}{\partial u}\D^{\alpha}\Bigr)
\Bigl(\frac{\partial}{\partial u}\D_{\alpha}\Bigr)
\end{equation}
where $\D^{\alpha}$ is the full ($\gamma$-dependent) radiation moment 
of the source $\J$.

It follows from the properties of ${\hat I}(x)$ above that, for the
calculation of (7.33) at $\I$, one needs to know the behaviour of 
$\log(-\Box)$ with a scalar test function that behaves at $\I$ like
\begin{equation}
X\|=\frac{1}{r^2}A(u,\phi)
\end{equation}
and at ${\rm i}^+$ like
\begin{equation}
X\biggl|_{\mbox{i}^+}=\frac{\gamma(1-\gamma^2)}{r}Q(\gamma,\phi)
\;\; ,\quad Q(1,\phi)\ne 0
\end{equation}
where $A(u,\phi)$ and $Q(\gamma,\phi)$ are some coefficients. The needed
result is obtained in Appendix C:
\begin{equation}
-\log(-\Box)X\|=2\frac{A(u,\phi)}{r^2}(\log r +{\bf c})+
\frac{B(u,\phi)}{r^2}+O\Bigl(\frac{\log r}{r^3}\Bigr)
\end{equation}
and
\begin{eqnarray}
\sphere B(u,\phi)\biggl|_{u\to\infty}&=&u\sphere\biggl[-6Q(1,\phi)+
4\int\limits_0^1 d\gamma\,\frac{\gamma^2}{1-\gamma^2}
\Bigl(Q(\gamma,\phi)-Q(1,\phi)\Bigr)\biggr]\nonumber \\
&&{}+O(\log u)\quad .
\end{eqnarray}

The behaviour of the function (7.33) at $\I$ is obtained by substituting
(7.36) for (7.39) and using (7.41) and (7.37). Summarizing the
calculation above one has
\begin{equation}
\T^{\mu\nu}(3)\|=\frac{1}{r^2}\nabla^{\mu}u\nabla^{\nu}u\;
\Bigl(-\frac{\partial\E(3)}{\partial u}\Bigr)+O\Bigl(\frac{1}{r^3}\Bigr)
\quad ,
\end{equation}
\begin{eqnarray}
-\frac{\partial\E(3)}{\partial u}&=&\frac{1}{(4\pi)^2}\,\tr\;
\biggl[-\frac{1}{6}(\log r +\log 2 +\frac{3}{2})
\Bigl(\frac{\partial}{\partial u}\d^{\alpha}\Bigr)
\Bigl(\frac{\partial}{\partial u}\d{}_{\alpha}\Bigr)\nonumber \\
&&{}+\frac{1}{6}\frac{\partial}{\partial u}
\int\limits^u_{-\infty}d\tau\,\log(u-\tau)
\Bigl(\frac{\partial}{\partial\tau}\d^{\alpha}(\tau)\Bigr)
\Bigl(\frac{\partial}{\partial\tau}\d{}_{\alpha}(\tau)\Bigr)\nonumber \\
&&{}-\frac{\partial^2}{\partial u^2}{\hat B}(u,\phi)+
\mbox{Q.N.}\biggr]\quad .
\end{eqnarray}
The contribution of the latter total-derivative term to the
radiation energy
\begin{equation}
\int\limits^{\infty}_{-\infty}du\sphere\Bigl(
-\frac{\partial\E(3)}{\partial u}\Bigr)
\end{equation}
is
\begin{equation}
-\frac{1}{(4\pi)^2}\,\tr\,\sphere\frac{\partial}{\partial u}
{\hat B}(u,\phi)\biggl|_{u\to\infty}\quad .
\end{equation}
Substituting (7.38) for (7.40) and using (7.42) one obtains
\begin{eqnarray}
-\sphere\frac{\partial}{\partial u}{\hat B}(u,\phi)\biggl|_{u\to\infty}
=\int\limits^{\infty}_{-\infty}du\sphere\Biggl\{\frac{1}{2}
\Bigl(\frac{\partial}{\partial u}\d^{\alpha}\Bigr)
\Bigl(\frac{\partial}{\partial u}\d{}_{\alpha}\Bigr)\hspace{2cm}\nonumber \\
{}-\frac{1}{3}\int\limits_0^1 d\gamma\,\frac{\gamma^2}{1-\gamma^2}\biggl[
\Bigl(\frac{\partial}{\partial u}\D^{\alpha}\Bigr)
\Bigl(\frac{\partial}{\partial u}\D_{\alpha}\Bigr)-
\Bigl(\frac{\partial}{\partial u}\d^{\alpha}\Bigr)
\Bigl(\frac{\partial}{\partial u}\d{}_{\alpha}\Bigr)\biggr]\Biggr\}\quad .
\end{eqnarray}
In this way the result (2.23) emerges.

}

%% file: arttxt08.tex
{\renewcommand{\theequation}{8.\arabic{equation}}

\begin{center}
\section{\bf    Other models}
\end{center}

$$ $$

The results for quantum-field models other than the standard loop
can be obtained by combining the results for the standard loop [3].
However, the results for the standard loop should then be known
in full, including the contributions of the potential $\P$. The
contribution of the potential to the vacuum energy flux is given
in expression (1.21) but this expression implies that the potential
is regular at $\I$:
\begin{equation}
\P\|=O\Bigl(\frac{1}{r^3}\Bigr)\quad .
\end{equation}
If condition (8.1) does not hold, the contribution of the potential
should be calculated anew. For the starting point one may take the 
expression for $\S(2)$ with the form factors in the high-frequency
approximation [2]
\begin{eqnarray}
\S(2)=\frac{1}{2(4\pi)^2}\integral\,\tr\,\biggl(-\frac{1}{2}
\P\log\Bigl(-\frac{\Box}{m^2}\Bigr)\P-\frac{1}{12}
\R_{\mu\nu}\log\Bigl(-\frac{\Box}{m^2}\Bigr)\R^{\mu\nu}\nonumber \\
{}+\mbox{const.}\:\P\P +\mbox{const.}\:\R_{\mu\nu}\R^{\mu\nu}\biggr)
\end{eqnarray}
and the expression for $\T^{\mu\nu}(3)\Bigl|_{\I}$ calculated in [2] up
to terms $O(\Box^0)$ in the argument $\Box$ of the external line:
\begin{equation}
\T^{\mu\nu}(3)\|=\frac{1}{(4\pi)^2}\nabla^{\mu}\nabla^{\nu}\;\tr\;
\log(-\Box){\hat I}(x)+O(\Box^0)\quad ,
\end{equation}
\begin{eqnarray}
{\hat I}(x)&=&\frac{1}{2}\Bigl(F_{11}(\Box_2 ,\Box_3 )-
\frac{1}{3}F_{00}(\Box_2 ,\Box_3 )\Bigr)\P_2\P_3\nonumber \\
&&{}-\frac{1}{6}\nabla_{\alpha}\nabla_{\beta}\;\frac{1}{\Box_2\Box_3}
\Bigl(F_{22}(\Box_2 ,\Box_3 )-2F_{11}(\Box_2 ,\Box_3 )\Bigr)
\J_2^{\;\alpha}\J_3^{\;\beta}\quad .
\end{eqnarray}
Of the missing terms $O(\Box^0)$, the important ones can easily be
restored. These are the terms in $L_{mn}$. Since the $\log(-\Box)$
in (8.3) originates from the expansion of the third-order form
factors (Appendix B), each $F_{mn}$ in (8.4) should be accompanied
by the respective $L_{mn}$ to form the combination (7.4). With the
terms in $L_{mn}$ added, expression (8.3) becomes analogous to (7.33).
The remaining terms $O(\Box^0)$ and the unspecified constants in (8.2) 
contribute only to a numerical renormalization of the news function
(Sec. 7).

For the first example, consider the spinor QED. The effective action 
generated by the fermion loop in this model is $(-1)$ times the action
for the standard loop with
\begin{equation}
\P=\frac{1}{2}\gamma^{\mu}\gamma^{\nu}\R_{\mu\nu}\quad ,
\quad \gamma^{\mu}\gamma^{\nu}+\gamma^{\nu}\gamma^{\mu}=2g^{\mu\nu}{\hat 1}
\end{equation}
and
\begin{equation}
\R_{\mu\nu}=-{\rm i}qF_{\mu\nu}{\hat 1}\quad ,\quad \tr\;{\hat 1}=4
\end{equation}
where $F_{\mu\nu}$ is the Maxwell tensor, and $q$ is the electron's
charge.

Since the potential (8.5) is singular at $\I$, one has to resort to
Eqs. (8.2)-(8.4). One obtains
\begin{equation}
\tr\;\P_2\P_3=-\frac{1}{2}\tr\;\R_2^{\;\mu\nu}\R_{3\mu\nu}
\end{equation}
which is valid with any insertion of the form $f(\nabla_2 ,\nabla_3 )$.
Then, by (6.7),
\begin{equation}
\tr\;\P_2\P_3=\tr\;\biggl[-\frac{1}{2}\;\Box\Bigl(\frac{1}{\Box_2\Box_3}
\J_2^{\;\alpha}\J_{3\alpha}\Bigr)+\nabla_{\alpha}\nabla_{\beta}
\Bigl(\frac{1}{\Box_2\Box_3}\J_2^{\;\alpha}\J_3^{\;\beta}\Bigr)+
\frac{1}{2}\Bigl(\frac{1}{\Box_2}\J_2^{\;\alpha}\J_{3\alpha}+
\J_2^{\;\alpha}\frac{1}{\Box_3}\J_{3\alpha}\Bigr)\biggr]\; .
\end{equation}
When this expression is inserted in (8.4)-(8.3), the contribution of the
first term vanishes at $\I$ because of the presence of the overall $\Box$ 
(see  Appendix B), and the contribution of the second term is pure
quantum noise because it has the same structure as the $\J\J$ term in
(8.4) but with no form factor $F_{22}$ \footnote{In each sum of $F_{nn}$
in (8.4), only the $F_{nn}$ with the highest $n$ is to be retained since 
the junior $F_{nn}$ contribute only to the quantum noise [2].}. As a
result, one is left with
\begin{equation}
{\hat I}(x)=\frac{1}{2}\;\frac{1}{\Box_3}\;F_{11}(\Box_2 ,\Box_3 )
\J_2^{\;\alpha}\J_{3\alpha}
-\frac{1}{6}\nabla_{\alpha}\nabla_{\beta}\;\frac{1}{\Box_2\Box_3}\;
F_{22}(\Box_2 ,\Box_3 )\J_2^{\;\alpha}\J_3^{\;\beta}+
\mbox{Q.N.}
\end{equation}
where the first term is the contribution of the potential.

For ${\hat I}(x)$ in (8.9) the technique of Ref. [2] yields straight
away
\begin{equation}
{\hat I}(x)\|=-\frac{1}{3r^2}\int\limits^u_{-\infty}d\tau\,(u-\tau)
\Bigl(\frac{\partial}{\partial\tau}\d^{\alpha}(\tau)\Bigr)
\Bigl(\frac{\partial}{\partial\tau}\d{}_{\alpha}(\tau)\Bigr)
\end{equation}
which is $(-2)$ times the expression (7.36), and
\begin{equation}
{\hat I}(x)\biggl|_{\mbox{i}^+}=-\frac{\gamma(1-\gamma^2)}{6r}
\int\limits^{\infty}_{-\infty}du\,
\Bigl(\frac{\partial}{\partial u}\D^{\alpha}\Bigr)
\Bigl(\frac{\partial}{\partial u}\D_{\alpha}\Bigr)
\end{equation}
which is $(-2)$ times the expression (7.38). Besides, there is the
overall $(-1)$ appropriate for fermions. It follows that
$\partial\E(3)/\partial u$  for QED is {\it twice} the result for
the standard loop. On the other hand, inserting (8.7) in (8.2) and
changing the overall sign, one finds that $\S(2)$ for QED is also
{\it twice} the result for the standard loop. Thus, up to a
numerical addition to the renormalization of the news function,
all the results for QED are obtained by doubling the respective results 
for the standard loop and making the substitution (8.6). Note that, since 
the balance between $\S(2)$ and $\S(3)$ is maintained, the fact of
doubling can be read just from the $\beta$-function.

Both the standard loop and the spinor QED have the "zero-charge" [11]
sign of the static vacuum polarization. It is interesting to see
what will be the results in the case of the "asymptotically free" sign.
For that, consider creation of the Yang-Mills quanta in the external
Yang-Mills field. In this consideration, it is convenient to refer
to the standard loop with the commutator curvature
\begin{equation}
\R_{\mu\nu}={\cal R}^a_{b\,\mu\nu}=C^a_{fb}F^f_{\mu\nu}
\end{equation}
where $C^a_{fb}$ are the group structure constants, and $F^f_{\mu\nu}$
is the strength of the external Yang-Mills field.

In the minimal [3] gauge, the effective action generated by the ghost loop
is $(-2)$ times the action for the standard loop with $\P=0$ and
$\R_{\mu\nu}$ in (8.12).

The quantities pertaining to the loop of the gauge field will be 
distinguished with a tilde and expressed through (8.12). The loop
of the gauge field is the standard loop with~[3]
\begin{eqnarray}
\widetilde{P}&=&P^{(a\alpha)}_{\;(b\beta)}=-2{\cal R}^a_{b\,\alpha\gamma}\:
g^{\gamma\beta}\quad , \\
\widetilde{{\cal R}}_{\mu\nu}&=&{\cal R}^{(a\alpha)}_{\:(b\beta)\,\mu\nu}
={\cal R}^a_{b\,\mu\nu}\:\delta^{\beta}_{\alpha}\quad .
\end{eqnarray}
Hence, in terms of the $\R_{\mu\nu}$ in (8.12),
\begin{eqnarray}
\tr\;\widetilde{P}_2\widetilde{P}_3&=&-4\,\tr\;\R_2^{\;\mu\nu}\R_{3\mu\nu}
\quad , \\
\tr\;\widetilde{{\cal R}}_2^{\;\mu\nu}\widetilde{{\cal R}}_{3\mu\nu}&=&
4\,\tr\;\R_2^{\;\mu\nu}\R_{3\mu\nu}\quad .
\end{eqnarray}
Relation (8.15) differs from (8.7) only in the coefficient. Therefore,
the calculation of the contribution of the potential repeats literally
the one above; only the result should be multiplied by 8. The contribution
of the potential $\widetilde{P}$ to $\partial\E(3)/\partial u$ is then
$(-24)$ times the result for the standard loop. The contribution of
$\widetilde{{\cal R}}_{\mu\nu}$ is, by (8.16), 4 times the result for the
standard loop. Since the contribution of ghosts is $(-2)$ times the
result for the standard loop, the grand total is $(-22)$ times the
result for the standard loop. The total action $\S(2)$ for the
Yang-Mills field is also $(-22)$ times the result for the standard loop
as follows immediately from inserting (8.15) and (8.16) in (8.2) and
adding the ghost contribution.

Thus, also for the Yang-Mills coupling, all vacuum fluxes are multiples  
of the respective fluxes for the standard loop (with the substitution (8.12)),
and the multiplicity is $(-22)$ in accord with the $\beta$-function
but the price for the asymptotic freedom is that the radiation energy is 
{\it negative}. However, having taken off the head they don't weep for
the hair [12]. Because the Yang-Mills quanta are exactly massless,
a source of the Yang-Mills field would cause initially an infinite
static polarization. The Yang-Mills charge is unobservable at infinity.

}

%% file: arttxt09.tex
\begin{center}
\section*{\bf Acknowledgments}
\end{center}

$$ $$

This work is dedicated to the memory of David Kirzhnits whose
understanding of physics enraptured, and whose paper [11] served
as a source of inspiration for the present author. The work was
supported in part by the Russian Foundation for Fundamental
Research Grant 96-02-16295 and INTAS Grant 93-493-ext.

%% file: arttxt10.tex
{\renewcommand{\theequation}{A.\arabic{equation}}

\section{\bf  The one-loop form factors}

$$ $$

The basic building element for all one-loop form factors [9] is the
operator\footnote{All operator functions are originally defined in
the Euclidean domain $\Box<0$.}
\begin{equation}
{\cal H}_q=\sqrt{\frac{2q}{\Box}}\;K_1\left(\sqrt{2q\Box}\right)
\;\; ,\quad q<0
\end{equation}
depending on the parameter $q$, with $K_1$ the order-1 Macdonald
function. By the properties of the Macdonald functions one has
also
\begin{equation}
\frac{d}{dq}{\cal H}_q=K_0\left(\sqrt{2q\Box}\right)
\end{equation}
and
\begin{equation}
2q\frac{d^2}{dq^2}{\cal H}_q=\Box\,{\cal H}_q\quad .
\end{equation}

Despite its scaring appearances the operator (A.1) has a simple
kernel. Its retarded kernel is
\begin{equation}
{\cal H}_qX(x)=\frac{1}{4\pi}\cint\delta\Bigl(\sigma(x,{\bar x})-q\Bigr)
X({\bar x})
\end{equation}
where $\sigma(x,{\bar x})$ is the world function [6], and the integration
is over the past sheet of the hyperboloid of equal geodetic distance from 
$x$. The derivation of (A.4) is based on the spectral representation
for the operator (A.1):
\begin{equation}
{\cal H}_q=-\frac{1}{\Box}-\sqrt{-2q}\;\int\limits_0^{\infty}d\mu\,
\frac{J_1\left(\mu\sqrt{-2q}\right)}{\mu^2-\Box}
\end{equation}
where $J_1$ is the Bessel function. Inserting in (A.5) the kernel of
the retarded resolvent [10]
\begin{equation}
\frac{1}{\mu^2-\Box}X(x)=\frac{1}{4\pi}\cint\biggl(\delta(\sigma)-
\theta(-\sigma)\frac{\mu J_1\left(\mu\sqrt{-2\sigma}\right)}{
\sqrt{-2\sigma}}\biggr)
{\bar X}\quad ,
\end{equation}
doing the spectral-mass integrations
\begin{eqnarray}
\int\limits_0^{\infty}d\mu\,J_1\bigl(\mu\sqrt{-2q}\bigr)
=\frac{1}{\sqrt{-2q}}\quad , \nonumber \\
\int\limits_0^{\infty}d\mu\,\mu\, J_1\bigl(\mu\sqrt{\mathstrut -2q}\bigr)
J_1\bigl(\mu\sqrt{\mathstrut -2\sigma}\bigr)
=\delta(\sigma -q)\quad ,
\end{eqnarray}
and using that
\begin{equation}
-\frac{1}{\Box}X(x)=\frac{1}{4\pi}\cint\delta(\sigma){\bar X}
\end{equation}
one obtains (A.4).

The kernel (A.4) was used in [2] without pointing out its relation 
to (A.1). This relation and the technique in [9] make it possible
to obtain the kernels of all one-loop form factors in the
expectation-value equations. Thus, for the second-order and
third-order form factors [9]
\begin{eqnarray}
F_{mn}(\Box_1 ,\Box_2 )&=&\left(\frac{\partial}{\partial j_1}\right)^m
\left(\frac{\partial}{\partial j_2}\right)^n
\left.\frac{\log(j_1\Box_1/j_2\Box_2)}{j_1\Box_1-j_2\Box_2}\right|_{
j_1=j_2=1}\quad ,\\
\Gamma_{kmn}(\Box_1 ,\Box_2 ,\Box_3 )&=&-\int_{\alpha>0}
d\alpha_1 d\alpha_2 d\alpha_3\,\delta (1-\alpha_1 -\alpha_2 -\alpha_3 )
\nonumber \\
&&\hspace{2cm}
{}\times\frac{\alpha_1{}^k\alpha_2{}^m\alpha_3{}^n}{\alpha_2\alpha_3\Box_1
+\alpha_1\alpha_3\Box_2 +\alpha_1\alpha_2\Box_3}
\end{eqnarray}
one has [9]
\begin{eqnarray}
F_{mn}(\Box_1 ,\Box_2 )&=&-2\int\limits^0_{-\infty}dq\,q^{m+n}
\frac{d^m}{dq^m}K_0 \left(\sqrt{2q\Box_1}\right)\,
\frac{d^n}{dq^n}K_0 \left(\sqrt{2q\Box_2}\right)\quad ,\\
\Gamma_{kmn}(\Box_1 ,\Box_2 ,\Box_3 )&=&\frac{4(-1)^{k+m+n}}{(k+m+n)!}
\int\limits^0_{-\infty}dq\,q^{k+m+n}\hspace{1cm}\nonumber \\
&&{}\times\frac{d^k}{dq^k}K_0 \left(\sqrt{2q\Box_1}\right)\,
\frac{d^m}{dq^m}K_0 \left(\sqrt{2q\Box_2}\right)\,
\frac{d^n}{dq^n}K_0 \left(\sqrt{2q\Box_3}\right)\quad\; 
\end{eqnarray}
and, therefore,
\begin{equation}
F_{mn}(\Box_1 ,\Box_2 )X_1 X_2(x)=-2\int\limits^0_{-\infty}dq\, q^{m+n}
\Bigl[\Bigl(\frac{d}{dq}\Bigr)^{m+1}{\cal H}_qX_1(x)\Bigr]
\Bigl[\Bigl(\frac{d}{dq}\Bigr)^{n+1}{\cal H}_qX_2(x)\Bigr]\quad ,
\end{equation}
\begin{eqnarray}
\Gamma_{kmn}(\Box_1 ,\Box_2 ,\Box_3 )X_1X_2X_3(x)=
\frac{4(-1)^{k+m+n}}{(k+m+n)!}
\int\limits^0_{-\infty}dq\, q^{k+m+n}\hspace{42mm}\nonumber \\
{}\times\Bigl[\Bigl(\frac{d}{dq}\Bigr)^{k+1}{\cal H}_qX_1(x)\Bigr]
\Bigl[\Bigl(\frac{d}{dq}\Bigr)^{m+1}{\cal H}_qX_2(x)\Bigr]
\Bigl[\Bigl(\frac{d}{dq}\Bigr)^{n+1}{\cal H}_qX_3(x)\Bigr]\quad
\end{eqnarray}
with ${\cal H}_qX(x)$ in (A.4).

Eq. (A.3) makes it possible to obtain easily the superpositions of
the kernels above with $1/\Box$. For example,
\begin{equation}
\frac{1}{\Box_2}F_{mn}(\Box_1 ,\Box_2 )X_1X_2 
=-\int\limits^0_{-\infty}dq\,q^{m+n}
\biggl(\frac{d^{m+1}}{dq^{m+1}}{\cal H}_qX_1\biggr)
\biggl(\frac{d^n}{dq^n}\int\limits^q_{-\infty}\frac{d{\bar q}}{{\bar q}}
{\cal H}_{\bar q}X_2\biggr)\quad ,
\end{equation}
\begin{equation}
\frac{1}{\Box_1\Box_2}F_{mn}(\Box_1 ,\Box_2 )X_1X_2 
=-\frac{1}{2}\int\limits^0_{-\infty}dq\,q^{m+n}
\biggl(\frac{d^m}{dq^m}\int\limits^q_{-\infty}\frac{d{\bar q}}{{\bar q}}
{\cal H}_{\bar q}X_1\biggr)
\biggl(\frac{d^n}{dq^n}\int\limits^q_{-\infty}\frac{d{\bar q}}{{\bar q}}
{\cal H}_{\bar q}X_2\biggr)
\end{equation}
which is valid including the cases $n=0$, $m=0$ (cf. [2]). Similarly for
the third-order form factors. The convergence of the integrals in $q$
at the upper limit is controlled by the behaviours
\begin{equation}
{\cal H}_q\biggl|_{q=0}=-\frac{1}{\Box}\quad ,\quad
\frac{d}{dq}{\cal H}_q\biggl|_{q\to 0}=O(\log q)
\end{equation}
following from (A.1), and the convergence at the lower limit should be 
provided by the properties of the test functions [2]. Eq. (A.3) can
also be obtained directly by acting with the operator $\Box$ on (A.4)
and neglecting the curvature in $\sigma$, $\Box\sigma =4+O[R]$.

One might have introduced the kernel even more elementary than (A.4):
\begin{equation}
\Theta_qX(x)=\frac{1}{4\pi}\cint\theta\Bigl(q-\sigma (x,{\bar x})\Bigr)
X({\bar x})\;\; ,\quad q<0
\end{equation}
\begin{equation}
\frac{d}{dq}\Theta_q={\cal H}_q\quad ,\quad 
\Theta_q\biggl|_{q=0}=\frac{2}{\Box^2}
\end{equation}
whence
\begin{equation}
\Theta_q=\frac{2q}{\Box}\;K_2\left(\sqrt{2q\Box}\right)\quad .
\end{equation}
The initial condition in (A.19) implies that
\begin{equation}
\frac{1}{\Box^2}X(x)=\frac{1}{8\pi}\cint\theta(-\sigma){\bar X}
\end{equation}
and
\begin{eqnarray}
\frac{1}{(4\pi)^2}\cint\delta\Bigl(\sigma(x,{\bar x})\Bigr)
\int\limits_{\mbox{\scriptsize past of}\;{\bar x}}d{\bar{\bar x}}\,
{\bar{\bar g}}^{1/2}\delta\Bigl(\sigma({\bar x},{\bar{\bar x}})\Bigr)
{\bar{\bar X}}\hspace{38mm}\nonumber \\
=\frac{1}{8\pi}\cint\theta\Bigl(-\sigma
(x,{\bar x})\Bigr){\bar X}\quad .
\end{eqnarray}
Eq. (A.21) can be obtained by acting with the operator $\Box$ on
(A.18) and using (A.8). It is also a limiting case of the formula [2]
\begin{equation}
\frac{1}{(m^2-\Box)^2}X(x)=\frac{1}{8\pi}\cint\theta(-\sigma)
J_0\left(m\sqrt{-2\sigma}\right){\bar X}
\end{equation}
for the massive operator. The kernels (A.18) and (A.21) do not decrease
at the future infinity and can be used for a direct determination of the
moments. Thus,
\begin{equation}
\frac{\partial}{\partial u}\;\biggl(\frac{1}{\Box^2}X\|\biggr)=
\frac{1}{2}D_{\bf 1}(u,\phi|X)
\end{equation}
where the quantity on the right-hand side is the $D_{\bf 1}$ moment
of the source $X$.

}

%% file: arttxt11.tex
{\renewcommand{\theequation}{B.\arabic{equation}}

\section{\bf  The third-order form factors at \protect\(\protect\I\protect\)}

$$ $$

The form factors in the third-order action (2.3) are linear combinations
of the functions $\Gamma_{kmn}(\Box_1 ,\Box_2 ,\Box_3 )$ introduced
in Appendix A. The typical contribution of such a form factor to the
energy-momentum tensor at point $x$ has the form
\begin{equation}
\Gamma_{kmn}(\Box ,\Box_2 ,\Box_3 )X_2X_3(x)
\end{equation}
where the $X$'s are the commutator curvatures or their derivatives, and 
it is assumed that first $\Box_2$ acts on $X_2=X(x_2)$, and $\Box_3$
on $X_3=X(x_3)$ with subsequently making the points $x_2$ and $x_3$ 
coincident  with the observation point $x$, and next the first
argument $\Box$ of the form factor acts on the thus obtained function
of the observation point. This nonlocal structure corresponds to the
diagram in Fig. 1.

By the results in Appendix A, expression (B.1) can be represented
as follows:
\begin{equation}
\Gamma_{kmn}(\Box ,\Box_2 ,\Box_3 )X_2X_3(x)=
\frac{4(-1)^{k+m+n}}{(k+m+n)!}\int\limits^0_{-\infty}dq\,
\Biggl(q^k\frac{d^k}{dq^k}K_0 \left(\sqrt{2q\Box}\right)\Biggr)
{\cal F}(q,x)
\end{equation}
where the operator $\Box$ acts to the right on the function of $x$,
and this function is
\begin{equation}
{\cal F}(q,x)=q^{m+n}
\Bigl[\Bigl(\frac{d}{dq}\Bigr)^{m+1}{\cal H}_qX_2(x)\Bigr]
\Bigl[\Bigl(\frac{d}{dq}\Bigr)^{n+1}{\cal H}_qX_3(x)\Bigr]\quad .
\end{equation}

When the $X$'s are expressed like in Eq. (6.7) through the sources
$J$ {\it having compact spatial supports}, there occur two
essentially different cases. An example of the first case is
\begin{equation}
X_2X_3=J_2^{\;\alpha}J_3^{\;\beta}\quad ,
\end{equation}
and examples of the second case are
\begin{equation}
X_2X_3=\frac{1}{\Box_2}J_2^{\;\alpha}J_3^{\;\beta}
\quad\mbox{or}\quad J_2^{\;\alpha}\frac{1}{\Box_3}J_3^{\;\beta}
\quad\mbox{or}\quad\nabla_{\alpha}\nabla_{\beta}
\Bigl(\frac{1}{\Box_2\Box_3}J_2^{\;\alpha}J_3^{\;\beta}\Bigr)\quad .
\end{equation}
The difference between the two cases is in the behaviours of integrals 
with the function ${\cal F}(q,x)$ as $x\to\I$. These behaviours are readily 
obtained by the technique in Ref. [2] (see also Sec. 6 above). In the
first case one has
\begin{equation}
\int\limits^0_{-\infty}dq\,{\cal F}(q,x)\:\biggl|_{x\to\I}=O(r^{-3})\quad ,
\end{equation}
and in the second case
\begin{equation}
\int\limits^0_{-\infty}dq\,{\cal F}(q,x)\:\biggl|_{x\to\I}=O(r^{-2})\quad .
\end{equation}
The second case is our main concern here since the function (B.7) is
singular at $\I$.

Our present goal is obtaining the behaviour of the current (B.2) as
$x\to\I$. The principal assertion is that this behaviour is determined 
by the first few terms of 
the \mbox{small-$\Box$} expansion of the form factor
$\Gamma$ in the argument $\Box$ of the external line. For the proof
it suffices to consider two generic terms of the small-$\Box$
expansion of the function $K_0 \left(\sqrt{2q\Box}\right)$ in (B.2):
\begin{equation}
(q\Box)^p\quad\mbox{and}\quad 
(q\Box)^p\log q\Box\quad .
\end{equation}

It will be recalled that the behaviours (B.6) and (B.7) are obtained
by making the replacement (6.10) of the integration variable $q$.  
From the form of this replacement, it follows that in the case (B.7)
one has also
\begin{eqnarray}
\int\limits^0_{-\infty}dq\,q^p\,{\cal F}(q,x)\:\biggl|_{x\to\I}&=&
r^{p-2}a_p(u,\phi)\quad , \\
\int\limits^0_{-\infty}dq\,q^p\,\log (-q){\cal F}(q,x)\:\biggl|_{x\to\I}&=&
r^{p-2}\log r\; a_p(u,\phi)+O(r^{p-2})\quad ,
\end{eqnarray}
and then, by the result in Appendix C,
\begin{equation}
\log(-\Box)\left.\left(\int\limits^0_{-\infty}dq\,q^p\,{\cal F}(q,x)
\right)\:\right|_{x\to\I}=-r^{p-2}\log r\; a_p(u,\phi)+
O(r^{p-2})\;\; ,\quad p\ge 1
\end{equation}
with one and the same coefficient $a_p(u,\phi)$ in all the expressions
(B.9) to (B.11).

Moreover, using the following form of the operator $\Box$ at $\I$ [2]:
\begin{equation}
\Box\,X\|=-\frac{2}{r}\frac{\partial}{\partial u}\,X
-2\frac{\partial^2}{\partial u\partial r}\,X+
O\Bigl(\frac{1}{r^2}X\Bigr)\quad ,
\end{equation}
one obtains
\begin{equation}
\Box\,O(r^p)\|=O(r^{p-1})\;\; ,\quad p\ne -1
\end{equation}
and, in the exceptional case $p=-1$,
\begin{equation}
\Box\,O(r^{-1})\|=O(r^{-3})\quad .
\end{equation}
Owing to the latter fact, one has
\begin{equation}
\Box^p\,O(r^{p-2})\|=O\Bigl(\frac{1}{r^3}\Bigr)\;\; ,\quad p\ge 1\quad .
\end{equation}

The relations above make it possible to obtain the contributions of the
expansion terms (B.8) to (B.2):
\begin{equation}
\Box^p\,\int\limits^0_{-\infty}dq\,q^p\,{\cal F}(q,x)\:\biggl|_{x\to\I}
=O\Bigl(\frac{1}{r^3}\Bigr)\;\; ,\quad p\ge 1
\end{equation}
\begin{equation}
\Box^p\,\Biggl[\log (-\Box)\biggl(\int\limits^0_{-\infty}dq\,q^p\,
{\cal F}(q,x)\biggr)+\int\limits^0_{-\infty}dq\,q^p\,\log (-q)
{\cal F}(q,x)\Biggr]\:\Biggr|_{x\to\I}=
O\Bigl(\frac{1}{r^3}\Bigr)\;\; ,\quad p\ge 1\; .
\end{equation}
It follows that, in the case (B.7), the function of $\Box$ in (B.2) 
can be truncated as follows:
\begin{eqnarray}
q^k\frac{d^k}{dq^k}K_0\left(\sqrt{2q\Box}\right)&=&
\left\{
\begin{array}{ccc}
{\displaystyle \frac{1}{2}(-1)^k(k-1)!}
&,&{\displaystyle \;\;k>0}\nonumber \\
{\displaystyle -\frac{1}{2}\log \frac{q\Box}{2}-{\bf c}}
&,&{\displaystyle \;\;k=0}\nonumber \\
\end{array}
\right.
\nonumber \\
&+&\mbox{irrelevant terms}
\end{eqnarray}
where the irrelevant terms are the terms whose contributions to (B.2) 
are $O(1/r^3)$ at $\I$. By a similar analysis, in the case (B.6) this 
function can be truncated even more:
\begin{eqnarray}
q^k\frac{d^k}{dq^k}K_0\left(\sqrt{2q\Box}\right)&=&
\left\{
\begin{array}{ccc}
{\displaystyle 0}
&,&{\displaystyle \;\;k>0}\nonumber \\
{\displaystyle -\frac{1}{2}\log (-\Box)}
&,&{\displaystyle \;\;k=0}\nonumber \\
\end{array}
\right.
\nonumber \\
&+&\mbox{irrelevant terms}\quad ,
\end{eqnarray}
and one recovers the algorithm used in [2]. Thus the amendment needed 
in the case (B.7) as compared to (B.6) is retaining the terms
$O(\Box^0)$ of the form factors.

Besides the contributions of the form (B.1), the vacuum energy-momentum
tensor contains contributions in which the form factors
$\Gamma_{kmn}(\Box ,\Box_2 ,\Box_3 )$ are superposed with $1/\Box$ in
the argument of the external line [9]:
\begin{equation}
\frac{1}{\Box}\;\Gamma_{kmn}(\Box ,\Box_2 ,\Box_3 )X_2X_3(x)\quad .
\end{equation}
These contributions occur only at $k\ge 1$ [9] and only in the case
(B.6) \footnote{This fact is a matter of a direct calculation [9]
but it is also a necessary condition for 
the expectation-value spacetime 
to be asymptotically flat.}. By the same consideration as above,
the operator function in (B.2) can then be truncated as follows:
\par\smallskip
\begin{eqnarray}
\frac{1}{\Box}\,q^k\frac{d^k}{dq^k}K_0\left(\sqrt{2q\Box}\right)&=&
\frac{1}{2}(-1)^k(k-1)!\,\frac{1}{\Box}\nonumber \\
&+&
\left\{
\begin{array}{ccc}
{\displaystyle -\frac{1}{4}(-1)^k(k-2)!\,q}
&,&{\displaystyle \;\;k>1}\nonumber \\
{\displaystyle -\frac{1}{4}\Bigl(\log \frac{q\Box}{2}+2{\bf c}-1\Bigr)q}
&,&{\displaystyle \;\;k=1}\nonumber \\
\end{array}
\right.
\nonumber \\
&+&\mbox{irrelevant terms} \quad .
\end{eqnarray}

The effect of these truncations is that the third-order form factors
boil down to the second-order form factors. The latter are the
functions $F_{mn}$ and $L_{mn}$ introduced in Sec.~7, and similar
functions originating from expansion (B.21) and differing from
$F_{mn}$ and $L_{mn}$ by an extra power of $q$:
\begin{equation}
G_{mn}(\Box_2 ,\Box_3 )X_2X_3=-2\int\limits^0_{-\infty}dq\,q^{m+n+1}
\Bigl[\Bigl(\frac{d}{dq}\Bigr)^{m+1}{\cal H}_qX_2\Bigr]
\Bigl[\Bigl(\frac{d}{dq}\Bigr)^{n+1}{\cal H}_qX_3\Bigr]\quad ,
\end{equation}
\begin{equation}
M_{mn}(\Box_2 ,\Box_3 )X_2X_3=-2\int\limits^0_{-\infty}dq\,
\left(\log(-\frac{q}{2})+2{\bf c}\right)q^{m+n+1}
\Bigl[\Bigl(\frac{d}{dq}\Bigr)^{m+1}{\cal H}_qX_2\Bigr]
\Bigl[\Bigl(\frac{d}{dq}\Bigr)^{n+1}{\cal H}_qX_3\Bigr]\; .\;\;
\end{equation}
Using Eq. (A.3), the latter functions can be expressed through
$F_{mn}$ and $L_{mn}$:
\begin{eqnarray}
G_{mn}(\Box_2 ,\Box_3 )&=&\frac{2}{\Box_2}\Bigl(F_{m+2,n}(\Box_2 ,\Box_3 )
+(m+1)F_{m+1,n}(\Box_2 ,\Box_3 )\Bigr) \nonumber \\
&=&\frac{2}{\Box_3}\Bigl(F_{m,n+2}(\Box_2 ,\Box_3 )+
(n+1)F_{m,n+1}(\Box_2 ,\Box_3 )\Bigr)\quad ,
\end{eqnarray}
\begin{eqnarray}
M_{mn}(\Box_2 ,\Box_3 )&=&\frac{2}{\Box_2}\Bigl(L_{m+2,n}(\Box_2 ,\Box_3 )
+(m+1)L_{m+1,n}(\Box_2 ,\Box_3)\Bigr)\nonumber \\
&=&\frac{2}{\Box_3}\Bigl(L_{m,n+2}(\Box_2 ,\Box_3 )+
(n+1)L_{m,n+1}(\Box_2 ,\Box_3 )\Bigr)\quad .
\end{eqnarray}
Equivalence of the two forms in (B.24) follows from the identities
for $F_{mn}$ in Ref. [2]. Similar identities can be derived for
$L_{mn}$, $G_{mn}$ and $M_{mn}$. All of them are based on Eq. (A.3)
and the integration by parts in the $q$ integrals.

The consideration above can be summarized as follows. For the case
(B.4) one has
\begin{eqnarray}
\Gamma_{kmn}(\Box ,\Box_2 ,\Box_3 )J_2J_3\|&=&
O\Bigl(\frac{1}{r^3}\Bigr)\;\; ,\quad k>0 \\
\Gamma_{{\bf 0}mn}(\Box ,\Box_2 ,\Box_3 )J_2J_3\|&=&
\frac{(-1)^{m+n}}{(m+n)!}\log(-\Box)\Bigl(F_{mn}(\Box_2 ,\Box_3 )
J_2J_3\Bigr)+O\Bigl(\frac{1}{r^3}\Bigr)\; .\;\;
\end{eqnarray}
For all subcases in (B.5) one has
\begin{eqnarray}
\Gamma_{kmn}(\Box ,\Box_2 ,\Box_3 )X_2X_3\|&=&
-\frac{(-1)^{m+n}(k-1)!}{(k+m+n)!}F_{mn}(\Box_2 ,\Box_3 )X_2X_3+
O\Bigl(\frac{1}{r^3}\Bigr)\;\; , \\
&&\hspace{75mm}k>0\nonumber \\
\Gamma_{{\bf 0}mn}(\Box ,\Box_2 ,\Box_3 )X_2X_3\|&=&
\frac{(-1)^{m+n}}{(m+n)!}\biggl[\log(-\Box)\Bigl(F_{mn}(\Box_2 ,\Box_3 )
X_2X_3\Bigr)\nonumber \\
&&\hspace{25mm}{}+L_{mn}(\Box_2 ,\Box_3 )X_2X_3\biggr]+
O\Bigl(\frac{1}{r^3}\Bigr)\;\; .\quad\qquad
\end{eqnarray}
For the superpositions of $\Gamma$ with $1/\Box$ one has
\begin{eqnarray}
\frac{1}{\Box}\;\Gamma_{kmn}(\Box ,\Box_2 ,\Box_3 )J_2J_3\|&=&
-\frac{(-1)^{m+n}(k-1)!}{(k+m+n)!}\;\frac{1}{\Box}\Bigl(
F_{mn}(\Box_2 ,\Box_3 )J_2J_3\Bigr)\nonumber \\
&&{}+\frac{1}{2}\frac{(-1)^{m+n}(k-2)!}{(k+m+n)!}G_{mn}
(\Box_2 ,\Box_3 )J_2J_3+O\Bigl(\frac{1}{r^3}\Bigr)\;\; ,\\
&&\hspace{75mm}k>1\nonumber \\
\frac{1}{\Box}\;\Gamma_{{\bf 1}mn}(\Box ,\Box_2 ,\Box_3 )J_2J_3\|&=&
-\frac{(-1)^{m+n}}{(m+n+1)!}\;\frac{1}{\Box}\Bigl(F_{mn}
(\Box_2 ,\Box_3 )J_2J_3\Bigr)\nonumber \\
&&{}-\frac{1}{2}\frac{(-1)^{m+n}}{(m+n+1)!}\biggl[\log(-\Box)
\Bigl(G_{mn}(\Box_2 ,\Box_3 )J_2J_3\Bigr)\nonumber \\
&&\hspace{15mm}{}+M_{mn}(\Box_2 ,\Box_3 )J_2J_3-
G_{mn}(\Box_2 ,\Box_3 )J_2J_3\biggr]\quad\qquad
\nonumber \\
&&{}+O\Bigl(\frac{1}{r^3}\Bigr)\;\; .
\end{eqnarray}
The senior terms of the latter expressions, proportional to $1/\Box$,
cancel in the energy-momentum tensor [1]. 

Another useful relation:
\begin{equation}
\frac{\Box}{\Box_2\Box_3}\;\Gamma_{kmn}(\Box ,\Box_2 ,\Box_3 )
J_2J_3\|=O\Bigl(\frac{1}{r^3}\Bigr)
\end{equation}
is a consequence of Eq. (B.14).

Finally, the relations (B.26)-(B.32) remain unchanged when multiplied
by $\Box_2/\Box_3$ or $\Box_3/\Box_2$. Indeed, replacing the function
${\cal F}(q,x)$ with $(\Box_2/\Box_3){\cal F}(q,x)$ or
$(\Box_3/\Box_2){\cal F}(q,x)$ doesn't change the behaviours
(B.6) and (B.7). Using Eq. (A.3), the multiplier $\Box_2/\Box_3$
or $\Box_3/\Box_2$ can be absorbed in any second-order or
third-order form factor.

}

%% file: arttxt12.tex
{\renewcommand{\theequation}{C.\arabic{equation}}

\section{\bf  The operators \protect\(\protect\log(-\protect\Box)\protect\) 
and \protect\(1/\protect\Box\protect\) for
test functions singular at \protect\(\protect\I\protect\)}

$$ $$

The behaviour (4.3) of the resolvent is valid only for test functions
having compact spatial supports [10]. Therefore, the behaviours at
$\I$ of all form factors, used in [2], are also valid only under
conditions (1.43). For $\log(-\Box)$ this is the behaviour (3.25). In 
the general case, the support of the test function may conventionally
be divided into a compact domain and asymptotic domain. The
$1/r^2$ behaviour in (3.25) is a contribution of the compact domain.
Any behaviour of $\log(-\Box)X\|$ more singular than $1/r^2$ (call
it just singular) can only be a contribution of the asymptotic
domain, i.e. of $X\|$ itself. Similarly, the regular behaviour of
$(1/\Box)X\|$ is $1/r$ and is a contribution of the compact domain.
A key to obtaining the contributions of the asymptotic domain is
the fact that the null hyperplane reaches $\I$ at only one point
of the celestial sphere [2]. Therefore, the singular contributions
are always local in the angles although possibly nonlocal in time.
To see why they may be nonlocal in time recall that, when a point
tends to $\I$, one generator of its past light cone merges with
$\I$ entirely [2]. The retarded time ranges along this generator
to $-\infty$ whereas the whole generator is labelled by a single
value of the angles.

Let ${\cal L}(x,{\bar x})$ be the retarded kernel of $\log(-\Box)\;$,
\begin{equation}
\log(-\Box)X(x)=\int d{\bar x}\,{\bar g}^{1/2}\,{\cal L}(x,{\bar x})
{\bar X}\quad .
\end{equation}
By the argument above,
\begin{equation}
\log(-\Box)X(x)\biggl|_{\textstyle x=(u,\phi,r\to\infty)}=
\int d{\bar x}\,{\bar g}^{1/2}\,{\cal L}(x,{\bar x})
\biggl({\bar X}\biggl|_{\textstyle {\bar\phi}=\phi}\biggr)+
O\Bigl(\frac{1}{r^2}\Bigr)\quad,
\end{equation}
i.e. for obtaining the singular terms at $\I$, the test function can
be taken at the angles of the observation point, ${\bar\phi}=\phi$.
The angle integrations in $d{\bar x}$ can then be done explicitly,
and, as a result, the kernel becomes spherically symmetric. Thus,
for obtaining the singular contributions at $\I$, it suffices to
consider the spherically symmetric kernel.

The spherically symmetric kernel suffices for obtaining also the
regular contributions provided that $X$ is a scalar, and one needs 
only the integral of (C.2) over the 2-sphere. Indeed, to lowest
order in the curvature, the scalar kernel of $\log(-\Box)$ can
depend on the angles only through the arc length between the
points $\phi$ and ${\bar\phi}$ on the 2-sphere. Therefore,
\begin{equation}
\sphere \biggl(\log(-\Box)X\biggr)=\int d{\bar x}\,{\bar g}^{1/2}\,
{\cal L}(x,{\bar x})\biggl(\int d^2{\cal S}({\bar\phi})\,
{\bar X}\biggr)
\end{equation}
where the angle integrations in $d{\bar x}$ concern only the kernel
${\cal L}(x,{\bar x})$ and convert it into a spherically symmetric
kernel.

One case of the singular behaviour considered below is where
$X\|=O(1/r)$, and $\log(-\Box)X\|$ is needed up to the regular terms
$O(1/r^2)$. In this case Eq. (C.2) works. Another case is where
$X\|=O(1/r^2)$, and $\log(-\Box)X\|$ is needed including the
regular terms $1/r^2$. This case is more difficult but is encountered 
only in $\T^{\mu\nu}(3)$ (Sec. 7) where the limitations implied in
Eq. (C.3) are fulfilled. Therefore, in both cases one may use
the spherically symmetric kernel.

Below, $y$ is a point of the 2-dimensional Lorentzian section of a 
spherically symmetric spacetime, and $Y(y)$ is a test function restricted
to this section. The spherically symmetric retarded kernel of the
operator $\log(-\Box)$ is of the form [8,5]
\begin{eqnarray}
-\log\Bigl(-\frac{\Box}{m^2}\Bigr)Y(y)&=&
\frac{1}{r}\int\limits^0_{\infty}d{\bar r}\,\frac{{\bar r}}{{\bar r}+r}
{\bar Y}\1 \nonumber \\
&+&\frac{1}{r}\int\limits^r_0 d{\bar r}\,\log\bigl(m(r-{\bar r})\bigr)
\frac{d}{d{\bar r}}\Bigl({\bar r}{\bar Y}\2\Bigr) \nonumber \\
&+&\frac{1}{r}\int\limits^r_{\infty}d{\bar r}\,
\log\bigl(m({\bar r}-r)\bigr)\frac{d}{d{\bar r}}
\Bigl({\bar r}{\bar Y}\3\Bigr) \nonumber \\
&+&2{\bf c}Y(y)
\end{eqnarray}
where $r$ is the luminosity coordinate of the observation point $y$,
and the integrations are along the null pathes 1,2,3 shown in Fig. 2.
In (C.4), each of the pathes is parametrized with the luminosity
coordinate ${\bar r}$. The retarded time labelling the radial future
light cones and normalized in (1.11) will be denoted $u$ as above. 
In the coordinates $y=(u,r)$, ${\bar y}=({\bar u},{\bar r})$, and
with the curvature neglected, path 1 is ${\bar u}+2{\bar r}=u$,
path 2 is ${\bar u}=u$, and path 3 is 
${\bar u}+2{\bar r}=u+2r$.

The spherically symmetric retarded kernel of the operator $1/\Box$
is of the form
\begin{equation}
-\frac{1}{\Box}Y(y)=\frac{1}{2r}\int\limits_{\Omega}d^2{\bar y}\,
g^{1/2}({\bar y})\,{\bar r}{\bar Y}
\end{equation}
where $\Omega$ is the domain bounded by the pathes 1,2,3, and
$d^2{\bar y}\,g^{1/2}({\bar y})$ is the induced volume element.
Hence, in the coordinates $y=(u,r)$,
\begin{equation}
-\frac{\partial}{\partial u}\;\frac{1}{\Box}Y(u,r)=
\frac{1}{2r}\int\limits^0_{\infty}d{\bar r}\,{\bar r}{\bar Y}\1
+\frac{1}{2r}\int\limits^r_0d{\bar r}\,{\bar r}{\bar Y}\2
-\frac{1}{2r}\int\limits^r_{\infty}d{\bar r}\,{\bar r}{\bar Y}\3\quad .
\end{equation}

Denoting the contributions of the pathes 1,2,3 in (C.4)
${\cal P}_1$, ${\cal P}_2$, ${\cal P}_3$, one has
\begin{equation}
-\log\Bigl(-\frac{\Box}{m^2}\Bigr)Y(y)={\cal P}_1(y)+{\cal P}_2(y)+
{\cal P}_3(y)+2{\bf c}Y(y)\quad .
\end{equation}
The contribution of path 2 can be rewritten identically as follows:
\begin{equation}
{\cal P}_2(y)=\frac{1}{r^{n+1}}\int\limits^r_0d{\bar r}\,{\bar r}\,
\frac{{\bar r}^n-r^n}{{\bar r}-r}{\bar Y}\2
+(\log mr)Y(y)+\frac{1}{r^n}\int\limits^1_0d\xi\,\log (1-\xi)f(r\xi)
\end{equation}
with $n$ arbitrary, and
\begin{equation}
f({\bar r})\equiv \frac{d}{d{\bar r}}\Bigl({\bar r}^{n+1}{\bar Y}\2\Bigr)
\quad .
\end{equation}
If one chooses $n$ equal to the power of decrease of $Y$ at $\I$
\begin{equation}
Y(y)\|=\frac{A(u)}{r^n}\quad ,
\end{equation}
the last integral in (C.8) will have a finite limit:
\begin{equation}
\int\limits^1_0d\xi\,\log(1-\xi)f(r\xi)\biggl |_{r\to\infty}=-A(u)\quad .
\end{equation}
In this way one obtains for $n=1$
\begin{equation}
n=1\;\; ,\;\;\quad {\cal P}_2(y)\|=\frac{\log(mr)}{r}A(u)
+O\Bigl(\frac{\log r}{r^2}\Bigr)\hspace{58mm}
\end{equation}
(with no pure $1/r$ term), and for $n=2$
\begin{equation}
n=2\;\; ,\;\;\quad {\cal P}_2(y)\|=2\frac{\log(mr)}{r^2}A(u)-
\frac{1}{r^2}\int\limits_0^{\infty}d{\bar r}\,\log(m{\bar r})
\frac{d}{d{\bar r}}\Bigl({\bar r}^2{\bar Y}\2\Bigr)+
O\Bigl(\frac{\log r}{r^3}\Bigr)\quad .
\end{equation}

As $y\to\I$, path 3 shifts entirely to $\I$. Introducing the retarded
time ${\bar u}$ as a parameter along path 3, one can easily
calculate the limit
\begin{equation}
{\cal P}_3(y)\|=\int\limits^u_{-\infty}d{\bar u}\,
\log\Bigl(m\frac{u-{\bar u}}{2}\Bigr)\frac{d}{d{\bar u}}
\Bigl({\bar Y}\|\Bigr)\quad .
\end{equation}
Hence for the behaviour (C.10) one obtains
\begin{equation}
{\cal P}_3(y)\|=\frac{1}{r^n}\int\limits^u_{-\infty}d{\bar u}\,
\log\Bigl(m\frac{u-{\bar u}}{2}\Bigr)\frac{d}{d{\bar u}}A({\bar u})
+O\Bigl(\frac{1}{r^{n+1}}\Bigr)\quad .
\end{equation}

Finally, as $y\to\I$, path 1 remains fixed. Therefore, its contribution
is always regular:
\begin{equation}
{\cal P}_1(y)\|=\frac{1}{r^2}\int\limits^0_{\infty}d{\bar r}\,{\bar r}\,
{\bar Y}\1\quad .
\end{equation}

The contributions of the pathes 1,2,3 in (C.6) are considered similarly.

In the case $n=2$ above, the total result is
\begin{equation}
-\log(-\Box)Y(y)\|=2\frac{A(u)}{r^2}(\log r + {\bf c})
+\frac{B(u)}{r^2}+O\Bigl(\frac{\log r}{r^3}\Bigr)
\end{equation}
where
\begin{equation}
B(u)=B_1(u)+B_2(u)\quad ,
\end{equation}
\begin{eqnarray}
B_1(u)&=&\int\limits^0_{\infty}d{\bar r}\,{\bar r}\,{\bar Y}\1
-\int\limits_0^{\infty}d{\bar r}\,\log{\bar r}\,\frac{d}{d{\bar r}}
\Bigl({\bar r}^2{\bar Y}\2\Bigr)\quad ,\\
B_2(u)&=&\int\limits^u_{-\infty}d{\bar u}\,
\log\Bigl(\frac{u-{\bar u}}{2}\Bigr)\,\frac{d}{d{\bar u}}A({\bar u})\quad ,
\end{eqnarray}
and the next task is obtaining the behaviour of the coefficient
$B(u)$ as $u\to\infty$.

The analysis of the behaviour of $B_1(u)$ at late time essentially repeats
the one in [2]. The dominant contribution to this behaviour comes from
$Y(y)$ at the limit $y\to\mbox{i}^+$ which in the present case is the
limit $r\to\infty$ along the radial timelike geodesic that reaches the
future infinity with the energy $E=(1-\gamma^2)^{-1/2}$ per unit rest mass:
\begin{equation}
y\to\mbox{i}^+\;\; :\;\;\qquad u=\frac{1-\gamma}{\gamma}\,r\;\; ,
\quad r\to\infty \quad .
\end{equation}
The variables $\gamma$ and $r$ may be used as coordinates of the point $y$ :
\begin{equation}
Y(y)=Y(\gamma,r)\quad.
\end{equation}
Then the definition of the limit $\mbox{i}^+$ is
\begin{equation}
Y\biggl |_{\mbox{i}^+}=Y(\gamma,r\to\infty)\quad .
\end{equation}
Of interest is the following behaviour of $Y$ at $\mbox{i}^+$ (see Sec. 7):
\begin{equation}
Y\biggl |_{\mbox{i}^+}=\frac{\gamma(1-\gamma^2)}{r}Q(\gamma)\;\; ,
\quad Q(1)\ne 0
\end{equation}
where $Q(\gamma)$ is some regular function of $\gamma$.

The limits $\mbox{i}^+$ and $\I$ are related [2]. For an analytic function,
the sequence of limits $\mbox{i}^+$ and $\gamma\to 1$ coincides with the
future of $\I$. Hence, using (C.21), one obtains
\begin{equation}
\Bigl(Y\|\Bigr)_{u\to\infty}=\Bigl(Y\biggl |_{\mbox{i}^+}\Bigr)_{\gamma\to 1}
=\frac{2u}{r^2}Q(1)\quad .
\end{equation}
Therefore, the behaviour (C.24) implies a linear growth of the coefficient
in (C.10) at late time:
\begin{equation}
A(u)\biggl |_{u\to\infty}=2uQ(1)\quad .
\end{equation}

The late-time behaviour of $B_1(u)$ in (C.19) is obtained by introducing
$\gamma$ as an integration variable in both integrals and 
restricting\footnote{The integration
limits $0<\gamma<1$ emerge after one restricts the support of ${\bar Y}$
to the interior of some future light cone ${\bar u}=\mbox{const.}$ and
the exterior of some tube ${\bar r}=\mbox{const.}$ The complementary portions
of the support of ${\bar Y}$ contribute negligibly as $u\to\infty$ [2].}
both integrations to the interval $0<\gamma<1$. One obtains
\begin{eqnarray}
B_1(u)\biggl|_{u\to\infty}&=&-u\int\limits_0^1\frac{d\gamma}{(1+\gamma)^2}
h_1(\gamma,r=\frac{\gamma u}{1+\gamma}\to\infty)\nonumber\\
&&{}-u\int\limits_0^1\frac{d\gamma}{(1-\gamma)^2}
h_2(\gamma,r=\frac{\gamma u}{1-\gamma}\to\infty)
\end{eqnarray}
where
\begin{eqnarray}
h_1(\gamma,r)&=&rY(\gamma,r)\quad ,\\
h_2(\gamma,r)&=&\log r\;\Bigl(\frac{\partial}{\partial r}+
\frac{\gamma(1-\gamma)}{r}\frac{\partial}{\partial\gamma}\Bigr)
r^2Y(\gamma,r)\quad .
\end{eqnarray}
With the behaviour (C.24) of $Y$ at $\mbox{i}^+$ this yields the result
\begin{equation}
B_1(u)\biggl |_{u\to\infty}=-u\int\limits_0^1d\gamma\,\gamma\,
\frac{1-\gamma}{1+\gamma}Q(\gamma)
-u\int\limits_0^1d\gamma\,\Bigl(\log \frac{u\gamma}{1-\gamma}\Bigr)
\frac{\partial}{\partial\gamma}\Bigl(\gamma^2(1+\gamma)Q(\gamma)\Bigr)
+O(\log u)\quad .
\end{equation}
The integration by parts brings this expression to the final form
\begin{equation}
B_1(u)\biggl |_{u\to\infty}=-2u(\log u -\log 2 +2)Q(1)
+4u\int\limits_0^1d\gamma\,\frac{\gamma^2}{1-\gamma^2}
\Bigl(Q(\gamma)-Q(1)\Bigr)+O(\log u)
\end{equation}
in which the coefficient of the linear growth is convergent. In this
way the pole at $\gamma=1$ is eliminated (cf. Eq. (1.21) and the
discussion of this problem in [2]).

The behaviour of $B_2(u)$ at late time is obtained by rewriting Eq. (C.20)
identically as follows:
\begin{equation}
B_2(u)=\int\limits^{u_0}_{-\infty}d{\bar u}\,
\log\Bigl(\frac{u-{\bar u}}{2}\Bigr)\frac{d}{d{\bar u}}A({\bar u})
+\Bigl(\log\frac{u}{2}\Bigr)\Bigl(A(u)-A(u_0)\Bigr)
+u\int\limits^1_{u_0/u}d\xi\,\log(1-\xi)\,g(u\xi)
\end{equation}
where $u_0<u$, and
\begin{equation}
g({\bar u})\equiv\frac{d}{d{\bar u}}A({\bar u})\quad .
\end{equation}
As $u\to\infty$, the first term in (C.32) is $O(\log u)$, and the
remaining terms are determined by the behaviour (C.26). In this way
one obtains
\begin{equation}
B_2(u)\biggl |_{u\to\infty}=2u\Bigl(\log\frac{u}{2}-1\Bigr)Q(1)+O(\log u)
\quad .
\end{equation}
In the sum (C.18) the senior terms $u\log u$ cancel, and the final result is
\begin{equation}
B(u)\biggl |_{u\to\infty}=u\Bigl[-6Q(1)+4\int\limits_0^1d\gamma\,
\frac{\gamma^2}{1-\gamma^2}\Bigl(Q(\gamma)-Q(1)\Bigr)\Bigr]
+O(\log u)\quad .
\end{equation}

Taking into account Eqs. (C.2) and (C.3), one can summarize the calculations
above as follows. For any function $X(x)$ in four dimensions that behaves
at $\I$ like
\begin{equation}
X\|=\frac{A(u,\phi)}{r^n}\;\; ,\quad n<2
\end{equation}
one has
\begin{equation}
-\frac{1}{\Box}\,X\|=\frac{1}{2(2-n)}\frac{1}{r^{n-1}}
\int\limits^u_{-\infty}d\tau\,A(\tau,\phi)+\frac{{\cal O}}{r^{n-1}}
\end{equation}
and
\begin{eqnarray}
-\log\Bigl(-\frac{\Box}{m^2}\Bigr)\,X\|&=&\frac{A(u,\phi)}{r^n}
\biggl(\log mr+2{\bf c}-\log 2 -1 +\int\limits^1_0
\frac{d\xi}{\xi^{n-1}}\frac{1-\xi^n}{1-\xi}\biggr)\nonumber\\
&&{}+\frac{1}{r^n}\int\limits^u_{-\infty}d\tau\,
\log\bigl(m(u-\tau)\bigr)\frac{\partial}{\partial\tau}A(\tau,\phi)
+\frac{{\cal O}}{r^n}
\end{eqnarray}
where ${\cal O}\|=0$. For a function $X(x)$ that behaves at $\I$ like
\begin{equation}
X\|=\frac{A(u,\phi)}{r^2}
\end{equation}
one has
\begin{equation}
-\frac{1}{\Box}\,X\|=\frac{1}{2}\frac{\log r}{r}\int\limits^u_{-\infty}
d\tau\,A(\tau,\phi)+O\Bigl(\frac{1}{r}\Bigr)
\end{equation}
and
\begin{equation}
-\log(-\Box)\,X\|=2\frac{A(u,\phi)}{r^2}(\log r +{\bf c})
+\frac{B(u,\phi)}{r^2}+O\Bigl(\frac{\log r}{r^3}\Bigr)
\end{equation}
with some coefficient $B(u,\phi)$. If in the latter case the function
$X(x)$ is a scalar that behaves at $\mbox{i}^+$ like
\begin{equation}
X\biggl |_{\mbox{i}^+}=\frac{\gamma(1-\gamma^2)}{r}Q(\gamma,\phi)\quad ,
\end{equation}
then
\begin{eqnarray}
\sphere B(u,\phi)\biggl |_{u\to\infty}&=&u\sphere \Bigl[-6Q(1,\phi)
+4\int\limits_0^1d\gamma\,\frac{\gamma^2}{1-\gamma^2}
\Bigl(Q(\gamma,\phi)-Q(1,\phi)\Bigr)\Bigr]\nonumber\\
&&{}+O(\log u)\quad .
\end{eqnarray}
Note that the term with $\log r$ in (C.41) is doubled as compared to (C.38).

}

%% file: arttxt13.tex
\begin{center}
\section*{\bf References}
\end{center}

$$ $$

\begin{enumerate}
\item A.G. Mirzabekian and G.A. Vilkovisky, Phys. Rev. Lett.
75 (1995) 3974; Class. Quantum Grav. 12 (1995) 2173.
\item A.G. Mirzabekian and G.A. Vilkovisky, Phys. Lett. B 414
(1997) 123; {\it Particle creation in the effective action method},
gr-qc@xxx.lanl.gov/9803006 (to appear in Ann. Phys., 1998).
\item A.O. Barvinsky and G.A. Vilkovisky, Phys. Rep. 119 (1985) 1.
\item R.M. Wald, {\it General relativity} (Chicago U.P., Chicago, 1984).
\item A.G. Mirzabekian, Zh. Eksp. Teor. Fiz. 106 (1994) 5 [Engl. trans.:
JETP 79 (1994) 1].
\item B.S. DeWitt, {\it Dynamical theory of groups and fields} (Gordon
and Breach, New York, 1965).
\item H. Bondi, M.G.J. van der Burg, and A.W.K. Metzner, Proc. R. Soc.
A 269 (1962) 21.
\item G.A. Vilkovisky, CERN-TH.6392/92; Publ. Inst. Rech. Math. Avanc\'ee,
R.C.P. 25, vol. 43 (Strasbourg, 1992) p. 203.
\item A.O. Barvinsky, Yu.V. Gusev, G.A. Vilkovisky, and V.V. Zhytnikov,
{\it Covariant perturbation theory} (IV). {\it Third order in the
curvature} (U. Manitoba, Winnipeg, 1993) pp. 1-192; J. Math. Phys.
35 (1994) 3525; {\it ibid} p. 3543; Nucl. Phys. B 439 (1995) 561;
Class. Quantum Grav. 12 (1995) 2157.
\item A.G. Mirzabekian and G.A. Vilkovisky, Phys. Lett. B 317 (1993) 517.
\item D.A. Kirzhnits, in: {\it Quantum field theory and quantum
statistics}, vol. 1, eds. I.A. Batalin, C.J. Isham, and G.A. Vilkovisky
(Hilger, Bristol, 1987) p. 349.
\item Russian proverb.
\end{enumerate}

%% file: arttxt14.tex
\begin{center}
\section*{\bf Figure captions}
\end{center}

$$ $$

\begin{itemize}
\item[Fig.1.] The diagram for the contribution (B.1) to the
expectation-value current at point $x$. The argument $\Box$
of the form factor $\Gamma$ corresponds to the external line.
\item[Fig.2.] Penrose diagram for the Lorentzian section of
a spherically symmetric spacetime. The timelike line $r=0$ 
is the central geodesic. The union of pathes 1 and 2, and
path 3 are the two radial light rays that come to the
2-dimensional observation point~$y$. $\Omega$ is the domain
bounded by the pathes 1,2,3.
\end{itemize}

%% file: figures.tex
\begin{figure}[p]
\begin{picture}(360,450)
\put(81,300){
\newbox\onebox
\newdimen\onew
\font\onea=onea at 72.27truept
\setbox\onebox=\vbox{\hbox{%
\onea\char0\char1\char2\char3}}
\onew=\wd\onebox
\setbox\onebox=\hbox{\vbox{\hsize=\onew
\parskip=0pt\offinterlineskip\parindent0pt
\hbox{\onea\char0\char1\char2\char3}
\hbox{\onea\char4\char5\char6\char7}
\hbox{\onea\char8\char9\char10\char11}}}
\ifx\parbox\undefined
    \def\setone{\box\onebox}
\else
    \def\setone{\parbox{\wd\onebox}{\box\onebox}}
\fi
\setone}
\end{picture}
\begin{center}
{\LARGE Fig.1}
\end{center}
\end{figure}
\newpage
\begin{figure}[p]
\begin{picture}(360,450)
\put(96,330){
\newbox\twobox
\newdimen\twow
\font\twoa=twoa at 72.27truept
\setbox\twobox=\vbox{\hbox{%
\twoa\char0\char1\char2}}
\twow=\wd\twobox
\setbox\twobox=\hbox{\vbox{\hsize=\twow
\parskip=0pt\offinterlineskip\parindent0pt
\hbox{\twoa\char0\char1\char2}
\hbox{\twoa\char3\char4\char5}
\hbox{\twoa\char6\char7\char8}
\hbox{\twoa\char9\char10\char11}}}
\ifx\parbox\undefined
    \def\settwo{\box\twobox}
\else
    \def\settwo{\parbox{\wd\twobox}{\box\twobox}}
\fi
\settwo}
\end{picture}
\begin{center}
{\LARGE Fig.2}
\end{center}
\end{figure}